\documentclass[11pt]{article}

\usepackage[preprint]{acl}
\usepackage[most]{tcolorbox}
\usepackage{times}
\usepackage{tabularx}
\usepackage{latexsym}
\usepackage{booktabs}
\usepackage{multirow}
\usepackage{amsmath}
\usepackage{algorithm}
\usepackage{algorithmic}
\usepackage[T1]{fontenc}
\usepackage[utf8]{inputenc}

\usepackage{microtype}

\usepackage{inconsolata}
\usepackage{graphicx}

\title{LLM-FK: Multi-Agent LLM Reasoning for Foreign Key Detection in Large-Scale Complex Databases}

\author{
  \textbf{Zijian Tang\textsuperscript{1}},
  \textbf{Ying Zhang\textsuperscript{1}}\thanks{\ \ Corresponding author.},
  \textbf{Sibo Cai\textsuperscript{2}},
\\
  \textbf{Ruoxuan Wang\textsuperscript{1}}
\\
\\
  \textsuperscript{1}Peking University,
  \textsuperscript{2}The Open University of China
\\
\texttt{\{zjtang25, wangrx24\}@stu.pku.edu.cn, zhang.ying@pku.edu.cn} \\
\texttt{caisb@ouchn.edu.cn}
}

\begin{document}
\maketitle

\begin{abstract}
Detecting missing foreign keys (FKs) requires accurately modeling semantic dependencies across database schemas, which conventional heuristic-based methods are fundamentally limited in capturing. We propose \textbf{LLM-FK}, the first fully automated multi-agent framework for FK detection, designed to address three core challenges that hinder naive LLM-based solutions in large-scale complex databases: combinatorial search space explosion, ambiguous inference under limited context, and global inconsistency arising from isolated local predictions. LLM-FK coordinates four specialized agents: 
a Profiler that decomposes the FK detection problem into the task of validating FK candidate column pairs and prunes the search space via a unique-key-driven schema decomposition strategy; 
an Interpreter that injects self-augmented domain knowledge; a Refiner that constructs compact structural representations and performs multi-perspective chain-of-thought reasoning; and a Verifier that enforces schema-wide consistency through a holistic conflict resolution strategy. Experiments on five benchmark datasets demonstrate that LLM-FK consistently achieves F1-scores above 93\%, surpassing existing baselines by 15\% on the large-scale MusicBrainz database, while reducing the candidate search space by two to three orders of magnitude without losing true FKs and maintaining robustness under challenging conditions like missing data. These results demonstrate the effectiveness and scalability of LLM-FK in real-world databases.
\end{abstract}

\section{Introduction}

Foreign key (FK) constraints define directional dependencies between a referencing column (or column set) in one table and a referenced column in another, thereby enforcing referential integrity in relational databases~\citep{melton2001sql}. These constraints play a foundational role in supporting essential downstream tasks, including schema reverse engineering, data integration, and semantic querying~\citep{wu2019discovering, li2023can}. However, in real-world databases, FK constraints are frequently missing due to schema evolution, legacy system migration, or intentional omission for performance reasons, which can undermine data reliability and semantic consistency~\citep{johnson2003database}. While manual FK identification may be feasible for small schemas, it quickly becomes impractical for large-scale databases characterized by numerous heterogeneous tables and complex data distributions~\citep{halevy2006principles, hai2023data}.

\begin{figure}
  \includegraphics[width=\columnwidth]{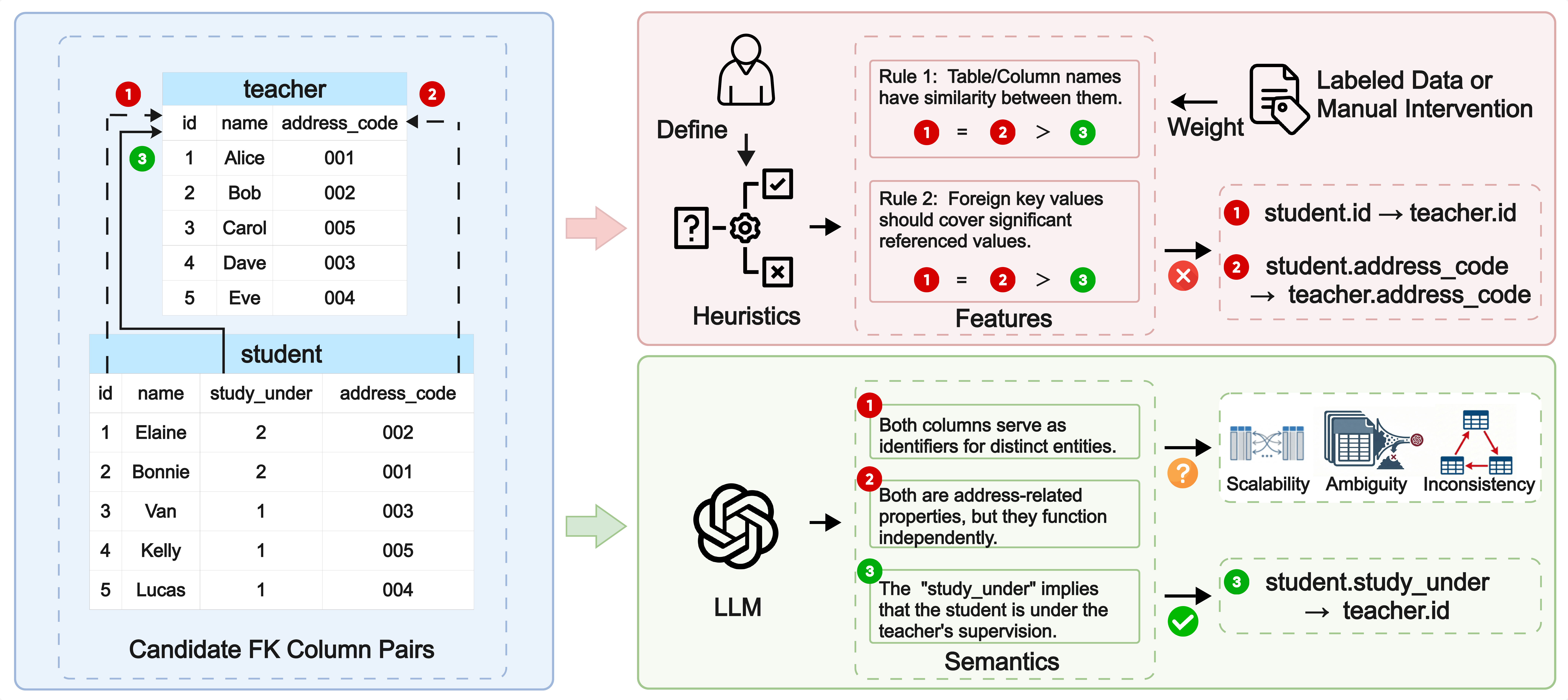}
\caption{Comparison of FK detection: heuristic limitations vs. LLM semantic reasoning.}
\label{fig:1}       % Give a unique label
\end{figure}

\begin{figure*}[t] 
  \centering
  \includegraphics[width=\linewidth]{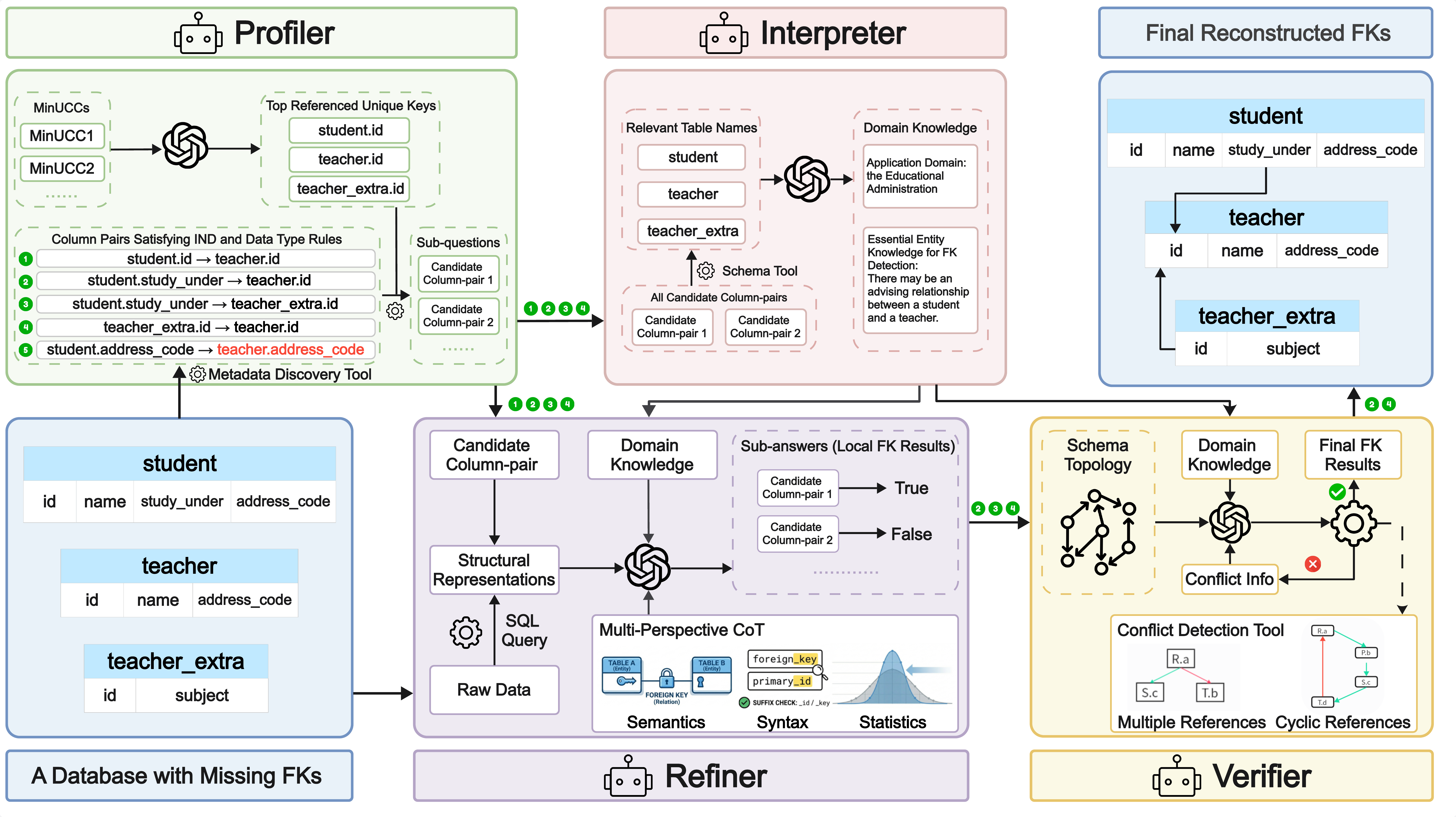}
  \caption{Overview of our LLM-FK framework. The pipeline orchestrates four specialized agents to achieve
        fully automated FK detection: the Profiler executes the Unique-Key-Driven Schema Decomposition Strategy; the
    Interpreter performs Self-Augmented Domain Knowledge Injection; the Refiner employs Multi-Perspective CoT Reasoning;
    and the Verifier applies the Holistic Conflict Resolution Strategy.}
  \label{fig:2}
\end{figure*}

% Existing automated FK detection methods predominantly rely on heuristic-based techniques that exploit syntactic or statistical signals, such as column name similarity and value inclusion ratios\citep{rostin2009machine, zhang2010multi, chen2014fast, jiang2020holistic, yan2020auto}. Although these heuristics are effective in well-structured or narrowly scoped schemas, they largely fail to capture the deeper semantic dependencies that FK constraints encode across tables \citep{lausen2007relational, kent2012database}. As a result, their performance is highly sensitive to data sparsity, unconventional naming conventions, and irregular schema designs, often necessitating manual supervision or high-quality labeled data to remain reliable. This dependence significantly limits their scalability and generalization to large, heterogeneous, or evolving databases.
Existing automated FK detection methods predominantly rely on heuristics that exploit syntactic or statistical signals, such as column name similarity and value inclusion ratios~\citep{rostin2009machine, chen2014fast, jiang2020holistic}. Although these heuristics are effective in well-structured or narrowly scoped schemas, they largely fail to capture the deeper semantic dependencies that fundamentally underpin FKs ~\citep{lausen2007relational, kent2012database}. 
As a result, their performance is highly sensitive to irregularities such as sparse data and unconventional naming patterns, and necessitates manual tuning or high-quality labeled data to remain reliable~\citep{zhang2010multi}, which significantly limits their scalability and generalizability to large, heterogeneous or evolving databases.

% As a result, their performance is highly sensitive to data sparsity, unconventional naming conventions, and irregular schema designs, often necessitating manual supervision or high-quality labeled data to remain reliable. 

Large Language Models (LLMs) offer a promising alternative by implicitly capturing semantic relationships that extend beyond surface-level heuristics; however, directly applying LLMs to large-scale FK detection introduces several fundamental challenges (see Figure \ref{fig:1}). First, \textbf{combinatorial search space explosion} severely limits scalability: as table and column counts increase, the number of candidate column pairs grows rapidly (e.g., approximately 4,000 pairs for only eight tables~\citep{poess2000new}), rendering exhaustive evaluation computationally infeasible. Second, \textbf{ambiguous inference under limited context} complicates accurate reasoning: finite context windows~\citep{ye2023large, chen2024tablerag} and performance degradation when processing raw tabular data ~\citep{cheng2022binding, chen2023large, sui2024tap4llm} prevent LLMs from jointly accessing complete schema structures and data distributions, forcing reliance on compressed or partial representations that may omit critical signals. 
Third, \textbf{global inconsistency arising from isolated local predictions} emerges because LLMs typically assess FKs in a local context independently, such as evaluating individual column pairs or table pairs in isolation, leading to globally conflicting decisions such as cyclic references that violate schema-level coherence.
% Third, \textbf{global inconsistency from local predictions} arises because LLMs typically assess each candidate column pair independently, leading to conflicting decisions such as cyclic references that violate schema-level coherence.

To overcome these limitations, we propose \textbf{LLM-FK}, a structured multi-agent framework for fully automated FK detection in large-scale complex databases, requiring neither manual intervention nor labeled data. LLM-FK decomposes schema-wide FK discovery into coordinated reasoning stages, with each stage explicitly addressing one of the aforementioned challenges. To mitigate \textbf{combinatorial search space explosion}, the Profiler employs a \textbf{Unique-Key-Driven Schema Decomposition Strategy}, leveraging LLMs to identify the most plausible referenced unique key for each potentially referenced table, and thereby decomposing the global detection task into focused validation problems over a pruned column pair candidate set. This constraint-aware decomposition dramatically reduces decision complexity and the search space while preserving recall of FKs. To address \textbf{ambiguous inference under limited context}, the Interpreter and Refiner jointly perform \textbf{Self-Augmented Domain Knowledge Injection} and \textbf{Multi-Perspective Chain-of-Thought (CoT) Reasoning}. Specifically, the Interpreter extracts implicit semantic cues from schema metadata and inferred domain knowledge, while the Refiner integrates this context with compact structural representations, enabling robust reasoning from syntactic, semantic, and statistical perspectives despite limited context. Finally, to ensure \textbf{global consistency from local predictions}, the Verifier applies a \textbf{Holistic Conflict Resolution Strategy} by constructing a schema-level dependency graph, detecting structural violations such as cyclic references, and performing targeted reasoning to resolve inconsistencies, thereby producing a coherent FK set aligned with relational semantics and overall schema topology.

We evaluate LLM-FK through extensive experiments on five public benchmark databases spanning diverse domains and scales. Overall, LLM-FK demonstrates consistent superiority in effectiveness, efficiency, and robustness. It achieves F1-scores exceeding 93\% on all benchmarks and outperforms both heuristic-based and LLM-based baselines by more than 15\% on MusicBrainz~\citep{swartz2002musicbrainz}, a complex real-world database containing over 300 tables. In terms of efficiency, LLM-FK reduces the candidate search space by two to three orders of magnitude without losing any true FKs. Furthermore, it maintains strong robustness under challenging conditions, including missing data and obscure semantic relationships. These results highlight the scalability and generalizability of LLM-FK for FK detection in large-scale complex databases.

\textbf{Our contributions} are summarized as follows:
\begin{itemize}
    \item We propose LLM-FK, the first multi-agent framework for fully automated FK detection in large-scale complex databases, overcoming the semantic limitations of existing approaches without requiring manual supervision or labeled data.

    \item We design four specialized agents that systematically address the core challenges of applying LLMs to FK detection: the Profiler reduces the search space via a Unique-Key-Driven Schema Decomposition Strategy; the Interpreter and Refiner enable deep and robust inference through Self-Augmented Domain Knowledge Injection and Multi-Perspective CoT Reasoning; and the Verifier enforces schema-wide coherence through a Holistic Conflict Resolution Strategy.

    \item Extensive experiments demonstrate that LLM-FK achieves state-of-the-art (SOTA)  accuracy, high efficiency, and strong robustness against complex, large-scale schemas.
\end{itemize}

\section{Related Work}

\subsection{Heuristic-based Methods for FK Detection}

Heuristic-based approaches constitute the dominant line of research for automated FK detection, relying on predefined syntactic and statistical signals extracted from database schemas and data instances. 
These methods can be broadly categorized into rule-weighted approaches and machine learning-based approaches.
Rule-weighted methods~\citep{zhang2010multi, chen2014fast, jiang2020holistic} design a collection of manually crafted heuristics—such as column name similarity and value inclusion ratios—and aggregate them through weighted scoring functions to estimate the plausibility of FK candidates.
Although effective in constrained settings, these approaches require labor-intensive manual tuning to determine appropriate heuristic weights and decision thresholds, which limits their adaptability across domains.

Machine learning-based methods attempt to alleviate manual tuning by learning decision functions from labeled examples, thereby automating candidate ranking or classification~\citep{rostin2009machine, yan2020auto}.
However, this paradigm introduces a strong dependency on high-quality labeled training data, which is expensive to obtain and often unavailable for large, real-world database schemas.
More fundamentally, because FKs encode semantic relationships rather than purely syntactic and statistical regularities, rigid heuristic signals inevitably break down in large-scale complex databases.
Irregularities such as sparse data, non-standard naming conventions, and heterogeneous schema designs frequently produce counterexamples that violate heuristic assumptions~\citep{zhang2010multi}, leading to degraded accuracy and limited generalization.
In contrast, our proposed LLM-FK leverages the semantic reasoning capabilities of LLMs to directly infer FK relationships, enabling automated and robust detection beyond the expressiveness of fixed heuristics.

\subsection{LLMs for Database Schema Analysis}

LLMs have recently shown strong potential in a range of database schema analysis tasks, including table annotation, entity matching, and join prediction~\citep{narayan2022can, fernandez2023large, jaimovitch2023can}.
These studies demonstrate that LLMs can effectively capture semantic information from schema metadata and natural language descriptions.
Despite this progress, systematic investigation of LLM-based database-wide FK detection remains largely unexplored.
Closely related works focus on pairwise join prediction. For instance, Chorus~\citep{kayali2023chorus} leverages Python code completion prompting to predict joinability between tables, but primarily targets ad-hoc analytical joins rather than formal FK constraints.
% Closely related works on pairwise join prediction, such as Chorus~\citep{kayali2023chorus}, leverage Python code completion to infer joinability, but primarily target ad-hoc analytical joins rather than formal FK constraints.
% Closely related works focus on pairwise join prediction. For instance, Chorus~\citep{kayali2023chorus} leverages Python code completion to predict pairwise joinability between tables, but primarily target ad-hoc analytical joins rather than formal FK constraints.
As a result, they do not explicitly model the semantic directionality, uniqueness requirements, or referential integrity guarantees that distinguish FKs from general joins. 

Moreover, MMQA preliminarily explores directly prompting LLMs to identify FKs between table pairs as a prerequisite for multi-table question answering~\citep{wu2025mmqa}; however, such generic LLM-based reasoning paradigms, including few-shot prompting and chain-of-thought reasoning, are difficult to apply naively to large-scale complex databases~\citep{brown2020language, wei2022chain}.
% Moreover, while MMQA\citep{wu2025mmqa} identifies FK detection as a prerequisite for multi-table QA, relying on generic LLM-based reasoning paradigms to identify foreign keys between table pairs is difficult to apply directly to large-scale complex databases~\citep{brown2020language, wei2022chain}.
% Moreover, generic LLM-based reasoning paradigms, including few-shot prompting and chain-of-thought reasoning, are difficult to apply directly to large-scale complex databases~\citep{brown2020language, wei2022chain}.
These approaches suffer from three fundamental limitations: combinatorial search space explosion when enumerating column pairs, ambiguous inference under limited context, and global inconsistency arising from isolated local predictions.
Without structural decomposition or coordination mechanisms, naive LLM applications struggle to scale while maintaining schema-level coherence.
To address these challenges, we propose LLM-FK, a multi-agent framework in which each agent is explicitly designed to resolve one of the limitations of applying LLMs to FK detection, enabling scalable, consistent, and semantically grounded inference.

\section{LLM-FK Approach}
\subsection{Problem Definition}
\label{sec:problem_definition}
FK detection aims to recover implicit referential constraints from a given relational database schema and its data instances.
Formally, given a relational database $\mathcal{D} = \{T_1, \dots, T_n\}$, each table $T_i$ is defined as a tuple $(\mathcal{C}_i, \mathcal{R}_i)$, where $\mathcal{C}_i$ denotes the set of columns (schema) and $\mathcal{R}_i$ denotes the finite set of data tuples (instance).
The objective of FK detection is to identify a set of elementary references $\Phi = \{(c_f, c_p) \mid c_f \in \mathcal{C}_i, c_p \in \mathcal{C}_j\}$, where each pair $(c_f, c_p)$ represents a single-column FK relationship in which the referencing column $c_f$ refers to the referenced column $c_p$.
This formulation generalizes to composite FKs~\citep{harrington2016relational}, which are modeled as combinations of multiple elementary references on distinct columns between the same source and target tables.
By reducing FK detection to the identification of elementary column-level references, this definition enables fine-grained reasoning while remaining compatible with higher-order relational constraints.

\subsection{Preliminaries: UCC and IND}
\label{sec:preliminaries}
\paragraph{Unique Column Combination (UCC) and Minimal UCC (MinUCC).} 
UCCs capture the notion of key-like column sets within a table.
Specifically, for a table $T_i$, a subset of columns $X \subseteq \mathcal{C}_i$ constitutes a \emph{UCC} if the projection of $T_i$ onto $X$ uniquely identifies every tuple.
Formally, for any two distinct tuples $t_u, t_v \in \mathcal{R}_i$, their projected values must satisfy $t_u[X] \neq t_v[X]$.
A \emph{MinUCC} further enforces minimality, requiring that no proper subset $X' \subset X$ preserves the uniqueness property.
MinUCCs are particularly important in FK detection, as they represent candidate keys and therefore constitute the only valid targets for FK references.

\paragraph{Inclusion Dependency (IND).} 
An \emph{IND}, denoted as $T_i[X] \subseteq T_j[Y]$, holds when the set of values appearing in column set $X \subseteq \mathcal{C}_i$ is contained within the set of values of $Y \subseteq \mathcal{C}_j$.
INDs provide a necessary structural condition for referential constraints: any FK from $X$ to $Y$ must satisfy the corresponding IND.
In this work, we focus exclusively on single-column INDs, which align with our elementary reference formulation and serve as a reliable basis for FK candidate pruning.

\subsection{Overview of LLM-FK}
\label{sec:overview}
We propose \textbf{LLM-FK}, a multi-agent framework for fully automated FK detection in large-scale complex databases.
As illustrated in Figure~\ref{fig:2}, LLM-FK consists of four specialized agents—\textbf{Profiler}, \textbf{Interpreter}, \textbf{Refiner}, and \textbf{Verifier}—each responsible for a distinct reasoning stage and equipped with a dedicated strategy.
Together, these agents decompose schema-wide FK discovery into coordinated, tractable sub-tasks that collectively address the limitations of naive LLM-based inference.

The \textbf{Profiler} initiates the workflow by decomposing the global detection task into column-pair validation problems and aggressively narrowing the candidate space through a Unique-Key-Driven Schema Decomposition Strategy.
The \textbf{Interpreter} then performs Self-Augmented Domain Knowledge Injection to expose implicit semantic signals embedded in the schema.
Building on this enriched context, the \textbf{Refiner} conducts deep inference using Multi-Perspective CoT Reasoning to evaluate each candidate FK.
Finally, the \textbf{Verifier} aggregates pairwise predictions and applies a Holistic Conflict Resolution Strategy to eliminate inconsistencies and ensure global schema coherence.
We detail each component and strategy in the following subsections, with additional implementation details provided in Appendix~\ref{More Implementation Details}.

\subsection{Unique-Key-Driven Decomposition}

The Profiler serves as the entry point of the LLM-FK pipeline, with the primary goal of optimizing LLM inputs for scalability and reliability.
It explicitly addresses two fundamental challenges: (1) the infeasibility of direct schema-wide FK detection due to limited LLM context windows and the risk of hallucinations, and (2) the combinatorial explosion of candidate FKs as schema size and complexity grow.
To this end, we introduce a \textbf{Unique-Key-Driven Schema Decomposition Strategy} that systematically transforms the global FK detection problem into a set of manageable column-level validation tasks while aggressively pruning structurally invalid candidates.

\paragraph{Schema Decomposition.}
We decompose FK detection into independent binary classification problems over directed column pairs.
Specifically, each candidate $(c_f, c_p)$ is evaluated in isolation, with the LLM determining whether $c_f$ functionally references $c_p$.
Compared to coarser decomposition granularities such as table-pair reasoning, column-level decomposition isolates the atomic unit of FK semantics.
This design keeps the sub-problem structure invariant to overall schema scale, focuses the LLM on a single directed hypothesis at one time, and substantially reduces decision complexity and hallucination risk.

\paragraph{Candidate Pruning.}
Although atomic decomposition simplifies individual decisions, it naively induces a combinatorial explosion in the number of column pairs.
To ensure computational feasibility, we employ a pruning strategy to aggressively refine the candidate set.
First, we apply IND-based validation and rule-based filters: INDs enforce necessary value inclusion constraints, while rule-based checks eliminate pairs with incompatible data types or semantically unsuitable types (e.g., Boolean or floating-point columns).
Second, we exploit the structural property that FKs must reference unique keys.
Instead of relying on primary keys (PKs)—which are frequently missing, incomplete, or overly composite—we identify MinUCCs as candidate referenced keys.
For each potentially referenced table, the LLM infers the most plausible referenced unique key from its MinUCC set.
As shown in Figure~\ref{fig:2}, we then discard any candidate column pair whose referenced column does not belong to the selected unique keys.
This process yields a compact yet comprehensive candidate set that preserves recall of true FKs while eliminating structurally invalid FK hypotheses.

\subsection{Knowledge-Augmented Reasoning}
\label{sec:knowledge_augmented_reasoning}

To accurately assess pruned candidate pairs under the constraints of finite LLM context windows, LLM-FK coordinates two agents—the Interpreter and the Refiner—to perform structured, knowledge-augmented reasoning.
The Interpreter focuses on Self-Augmented Domain Knowledge Injection to resolve semantic ambiguity, while the Refiner employs Multi-Perspective CoT Reasoning to conduct deep, evidence-based inference.

\paragraph{Self-Augmented Domain Knowledge Injection.}
The Interpreter equips the framework with a global semantic understanding of the database by analyzing the names of all tables involved in the candidate set.
We intentionally restrict this analysis to table names, as they capture core entity semantics while remaining sufficiently concise to fit within the LLM’s context budget.
By inferring the overarching application domain and salient entity concepts, the Interpreter bridges the gap between abstract schema symbols and real-world semantics.
Crucially, this process enables \emph{self-augmentation}~\citep{sui2024table}: schema-wide domain insights are injected into each reasoning task, reducing ambiguity when candidate column pairs are evaluated within an isolated local context.
The Interpreter performs this analysis once per database, ensuring consistent semantic grounding across all subsequent validations without redundant computation.

\paragraph{Multi-Perspective CoT Reasoning.}
Building on the injected domain knowledge, the Refiner implements a multi-perspective CoT framework that integrates complementary signals for robust inference.
To this end, the Refiner issues SQL queries to extract and serialize compact, high-density representations of relevant information, including:
(1) schema definitions (table and column names);
(2) statistical metadata (ordinal positions, data types, value ranges, cardinalities, row counts, and average text lengths);
(3) inter-column dependencies (value coverage ratios and table size ratios); and
(4) sample data instances (top-5 rows).
% This concise serialization avoids the noise and cost associated with full-table inputs while preserving discriminative signals.
% Crucially, by explicating these structural details, the Refiner provides sufficient evidence to disambiguate misleading surface features (e.g., numerical formats interpreted as strings), preventing premature misinterpretations common in raw data processing.
This concise serialization avoids the noise and cost associated with full-table inputs while preserving sufficient and discriminative evidence to prevent misinterpretations in raw data processing, such as ambiguous data types.
The Refiner then reasons from three complementary perspectives: a syntactic perspective based on naming conventions, a statistical perspective grounded in distributional regularities, and a semantic perspective focused on entity relationships.
By emulating expert human reasoning, this multi-perspective process enables cross-perspective compensation, allowing reliable signals to correct judgments when others are compromised, such as in obfuscated schemas or data-sparse tables.

\subsection{Holistic Conflict Resolution}
To ensure schema-wide structural consistency, the Verifier applies a Holistic Conflict Resolution Strategy that operates over aggregated pairwise predictions.
It constructs a schema topology graph encoding inferred FK relationships and then identifies structural violations arising from local decisions.
In particular, we focus on two common conflict types: \emph{multiple references}, where a single column references multiple distinct targets, and \emph{cyclic references}, where FK relationships form closed loops.

Conflict resolution proceeds in two stages.
First, for multiple-reference conflicts, the Verifier reevaluates competing candidates using the same knowledge-augmented reasoning framework described in Section~\ref{sec:knowledge_augmented_reasoning}, retaining only the most semantically and structurally plausible FK.
Second, for cyclic references, the Verifier employs an iterative elimination strategy that prioritizes the shortest cycles, which represent the most immediate violations of relational semantics.
Within each cycle, the least plausible reference is removed, and the process repeats until the graph converges to an acyclic and unambiguous state.
Crucially, the Refiner's semantic-based reasoning typically constrains the candidates into tightly clustered, semantically related subgraphs. Leveraging this localization, our approach yields two primary advantages. First, unlike methods that rely on isolated confidence scores, evaluating conflicting candidates collectively within these specific structures facilitates direct semantic comparison, accurately capturing intrinsic semantic hierarchical relationships. Second, it significantly reduces computational complexity; by confining conflict resolution to these highly localized immediate contexts (e.g., length-2 cyclic references), our approach intrinsically avoids the cascading complexity typically associated with global graph traversals.
Through this procedure, the Verifier guarantees that the final FK set is globally coherent with relational database principles.

\section{Experiments}
\subsection{Experimental Setup}
\label{Experiment Setup}
% \paragraph{Datasets and Evaluation Metrics.} We evaluate our framework on five widely adopted database benchmarks: TPC-H \cite{poess2000new}, Northwind(\textbf{NWind}) \cite{MSNorthwind}, TPC-E \cite{chen2011tpc}, AdventureWorks(\textbf{AdvWorks}) \cite{MSAdventureWorks}, and Musicbrainz(\textbf{MusicBz}) \cite{swartz2002musicbrainz}. 
% % As detailed in Table~\ref{tab:dataset_details}, 
% These datasets span diverse domains—including commerce, retail, finance, manufacturing, and entertainment—covering a broad spectrum of schema complexities and diverse real-world challenges. To assess the final FK detection quality, we employ standard Precision (P), Recall (R), and F1-score (F) \citep{goutte2005probabilistic}. These metrics are computed based on the candidate set generated by the Profiler, as this constitutes the effective search space for the task. Dataset details are provided in Appendix \ref{dataset_details}.

\paragraph{Datasets and Evaluation Metrics.}
We evaluate LLM-FK on five widely adopted relational database benchmarks: TPC-H~\citep{poess2000new}, Northwind (\textbf{NWind})~\citep{MSNorthwind}, TPC-E~\citep{chen2011tpc}, AdventureWorks (\textbf{AdvWorks})~\citep{MSAdventureWorks}, and MusicBrainz (\textbf{MusicBz})~\citep{swartz2002musicbrainz}.
These datasets span diverse application domains—including commerce, retail, finance, manufacturing, and entertainment—and exhibit substantial variation in schema scale, naming conventions, and data characteristics.
Together, they capture a broad spectrum of real-world challenges for FK detection, ranging from clean benchmark schemas to large-scale, heterogeneous databases. Detailed dataset statistics are reported in Appendix~\ref{dataset_details}.

To assess FK detection quality, we adopt standard Precision (P), Recall (R), and F1-score (F)~\citep{goutte2005probabilistic}.
All metrics are computed over the candidate set produced by the Profiler, which defines the effective search space for subsequent reasoning.
This evaluation protocol isolates the quality of FK inference from trivial candidates eliminated during early pruning.

% \paragraph{Baselines.} We conduct comprehensive experiments comparing our framework against seven baselines across three paradigms: (1) Rule-weighted Methods. MC-FK \citep{zhang2010multi} relies on pure data statistics assuming uniform distributions. Fast-FK \citep{chen2014fast} combines syntactic similarity with cardinality ratios. HOPF \citep{jiang2020holistic} leverages PK constraints to enhance detection performance. (2) Machine Learning-based Methods. ML-FK \citep{rostin2009machine} employs classifiers on basic features like value coverage. AutoSuggest \citep{yan2020auto} extends this by incorporating metadata such as ordinal positions. (3) LLM-based Methods. Chorus \citep{kayali2023chorus} performs join prediction via Python code completion. End-to-End queries LLMs to identify FKs for each table pair in a single pass. Few-Shot extends this strategy by incorporating explicit FK exemplars. CoT \citep{wei2022chain} employs chain-of-thought prompting to elicit step-by-step reasoning.

\begin{table*}[t]
\centering
% --- 调整区域开始 ---
% 1. 增大行高：建议设置为 1.2 到 1.3 之间 (原为 0.98)
\renewcommand{\arraystretch}{1.25}
% 2. 适度增大列间距：建议 4pt 到 5pt (原为 3pt)
% 注意：列数太多，如果此值过大，会导致 resizebox 把字体缩得太小
\setlength{\tabcolsep}{4pt} 
% --- 调整区域结束 ---

\resizebox{\textwidth}{!}{
\footnotesize
\begin{tabular}{l|l|ccc|ccc|ccc|ccc|ccc}
\hline
\multirow{2}{*}{\textbf{Category}} & \multirow{2}{*}{\textbf{Method}} & \multicolumn{3}{c|}{\textbf{TPC-H}} & \multicolumn{3}{c|}{\textbf{Northwind}} & \multicolumn{3}{c|}{\textbf{TPC-E}} & \multicolumn{3}{c|}{\textbf{AdventureWorks}} & \multicolumn{3}{c}{\textbf{MusicBrainz}} \\
\cline{3-17}
& & \textbf{P} & \textbf{R} & \textbf{F} & \textbf{P} & \textbf{R} & \textbf{F} & \textbf{P} & \textbf{R} & \textbf{F} & \textbf{P} & \textbf{R} & \textbf{F} & \textbf{P} & \textbf{R} & \textbf{F} \\
\hline
\multirow{3}{*}{Rule-weighted} 
& MC-FK & 0.56 & \textbf{1.00} & 0.72 & \underline{0.71} & \underline{0.91} & 0.80 & 0.32 & 0.78 & 0.46 & 0.46 & 0.59 & 0.52 & 0.51 & 0.81 & 0.62 \\
& Fast-FK & 0.50 & 0.78 & 0.61 & 0.67 & \underline{0.91} & 0.77 & \underline{0.82} & \underline{0.91} & \underline{0.86} & 0.54 & \underline{0.84} & 0.65 & 0.62 & 0.81 & 0.70 \\
& HOPF & \textbf{1.00} & 0.10 & 0.20 & 0.67 & 0.19 & 0.29 & 0.78 & 0.16 & 0.26 & 0.57 & 0.45 & 0.46 & \underline{0.75} & \underline{0.86} & \underline{0.80} \\
\hline
\multirow{2}{*}{ML-based} 
& ML-FK & 0.67 & 0.67 & 0.67 & 0.00 & 0.00 & 0.00 & 0.79 & 0.58 & 0.67 & \underline{0.77} & 0.22 & 0.34 & 0.35 & 0.49 & 0.41 \\
& AutoSuggest & 0.50 & 0.22 & 0.31 & 0.48 & \underline{0.91} & 0.62 & 0.45 & 0.64 & 0.53 & 0.24 & 0.54 & 0.33 & 0.30 & 0.23 & 0.26 \\
\hline
\multirow{4}{*}{LLM-based} 
& Chorus & 0.50 & 0.44 & 0.47 & 0.46 & 0.55 & 0.50 & 0.32 & 0.47 & 0.38 & 0.13 & 0.29 & 0.18 & 0.26 & 0.32 & 0.29 \\
& End-to-End & \underline{0.82} & \textbf{1.00} & \underline{0.90} & \textbf{1.00} & \textbf{1.00} & \textbf{1.00} & 0.57 & 0.87 & 0.69 & 0.72 & 0.75 & \underline{0.73} & 0.70 & 0.81 & 0.75 \\
& Few-Shot & 0.80 & \underline{0.89} & 0.84 & \textbf{1.00} & \underline{0.91} & \underline{0.95} & 0.46 & 0.71 & 0.56 & 0.63 & 0.68 & 0.66 & 0.70 & 0.70 & 0.70 \\
& CoT & \underline{0.82} & \textbf{1.00} & \underline{0.90} & \textbf{1.00} & \textbf{1.00} & \textbf{1.00} & 0.60 & 0.87 & 0.71 & 0.71 & 0.74 & 0.72 & 0.68 & 0.81 & 0.74 \\
\hline
\textbf{Ours} & \textbf{LLM-FK} & \textbf{1.00} & \textbf{1.00} & \textbf{1.00} & \textbf{1.00} & \textbf{1.00} & \textbf{1.00} & \textbf{0.91} & \textbf{0.96} & \textbf{0.93} & \textbf{0.93} & \textbf{0.96} & \textbf{0.94} & \textbf{0.91} & \textbf{1.00} & \textbf{0.95} \\
\hline
\end{tabular}
}
\caption{Performance comparison of FK detection across five datasets. \textbf{Bold} indicates the best performance, and \underline{underlined} indicates the second-best.}
\label{tab:overall_performance}
\end{table*}

\paragraph{Baselines.}
We compare LLM-FK against seven representative baselines spanning three methodological paradigms.
\emph{(1) Rule-weighted methods.}
MC-FK~\citep{zhang2010multi} relies exclusively on data statistics under uniform reference assumptions.
Fast-FK~\citep{chen2014fast} combines syntactic similarity with cardinality-based signals.
HOPF~\citep{jiang2020holistic} further exploits PK constraints to improve detection performance.
\emph{(2) Machine learning-based methods.}
ML-FK~\citep{rostin2009machine} trains classifiers on basic syntactic and statistical features.
AutoSuggest~\citep{yan2020auto} extends this approach by incorporating additional metadata, including ordinal positions.
\emph{(3) LLM-based methods.}
Chorus~\citep{kayali2023chorus} predicts joinability via Python code completion prompting.
End-to-End~\citep{wu2025mmqa} directly queries LLMs to identify FKs for each table pair in a single pass.
Few-Shot~\citep{brown2020language} augments this setting with explicit FK exemplars.
CoT~\citep{wei2022chain} applies chain-of-thought prompting to elicit step-by-step reasoning.
These baselines collectively represent the current SOTA across heuristic-based and LLM-driven paradigms.

% \paragraph{Implementation Details.}
% We employ \textbf{DeepSeek-R1} \citep{guo2025deepseek} with temperature 0 for reproducibility. To prevent data leakage, we mask database names; Additionally, empty columns are excluded during rule-based pruning in the Profier to ensure data-driven baselines efficacy. For further resources, Appendix~\ref{Reproducibility Statement} provides the reproducibility statement, while Appendix~\ref{sec:more_experiment} presents supplementary experiments with Qwen3-Max\citep{yang2025qwen3} and detailed example analyses.

\paragraph{Implementation Details.}
All LLM-based methods are implemented using \textbf{DeepSeek-R1}~\citep{guo2025deepseek} with the temperature set to 0 to ensure deterministic and reproducible outputs.
To prevent data leakage~\citep{balloccu2024leak}, database names are masked during inference.
In addition, empty columns are excluded during rule-based pruning in the Profiler to ensure fair comparison with data-driven baselines.
Further implementation details and reproducibility considerations are provided in Appendix~\ref{Reproducibility Statement}.
% Supplementary experiments using other LLMs, a detailed computational cost analysis, additional results, and qualitative case studies are presented in Appendix~\ref{sec:more_experiment}.
% Supplementary experiments using Qwen3-Max~\citep{yang2025qwen3}, along with qualitative case studies, are presented in Appendix~\ref{sec:more_experiment}.
Supplementary experiments—including evaluations on other LLMs, a detailed computational cost analysis, and other additional results—along with qualitative case studies, are presented in Appendix~\ref{sec:more_experiment}.

\subsection{Main Results}
\label{Main Results}

Table~\ref{tab:overall_performance} summarizes the overall FK detection performance.
The results reveal several key findings:
\textbf{First}, LLM-FK consistently achieves SOTA performance across all five datasets, outperforming the strongest baselines by substantial margins on complex schemas (e.g., +21\% F1-score on AdventureWorks and +15\% on MusicBrainz).
\textbf{Second}, LLM-FK exhibits markedly higher accuracy and generalizability than heuristic-based methods.
While the strongest heuristic-based baseline (Fast-FK) attains F1-scores ranging from 0.61 to 0.86—exhibiting large variance across datasets—LLM-FK maintains stable performance between 0.93 and 1.00.
This improvement stems from LLM-FK’s ability to reason about semantic intent, which is fundamental to FK design, rather than relying solely on brittle syntactic or statistical signals that are sensitive to irregularities.
\textbf{Third},  while generic LLM-based baselines degrade sharply on complex schemas like AdventureWorks and MusicBrainz, LLM-FK demonstrates superior stability. Specifically, our multi-agent collaboration framework decomposes FK detection into invariant, binary directed column-pair validation tasks to reduce decision complexity, executes deep reasoning despite limited context, and enforces global schema-wide consistency, thereby effectively overcoming the challenges of native LLM-based paradigms on large-scale complex databases.
% compared to generic LLM-based approaches, LLM-FK demonstrates substantially better stability on complex databases.
% Although these baselines perform competitively on simpler schemas, their accuracy degrades sharply on large and intricate benchmarks like AdventureWorks and MusicBrainz.
% By decomposing FK detection into invariant, binary column-pair validation tasks, LLM-FK limits reasoning scope, concentrates LLM attention, and effectively mitigates hallucinations.
Moreover, the poor performance of Chorus confirms that join prediction alone fails to capture the semantic directionality and integrity constraints intrinsic to FKs.
\textbf{Finally}, LLM-FK shows strong robustness under semantic opacity, as exemplified by TPC-E, which contains numerous ambiguous English abbreviations.
While generic LLM-based methods struggle in this setting, LLM-FK achieves an F1-score of 0.93, benefiting from Self-Augmented Domain Knowledge Injection and Multi-Perspective CoT Reasoning that jointly disambiguate local column pairs using global and structural cues.
Overall, these results demonstrate that LLM-FK provides accurate, generalizable, and robust FK detection for large-scale complex databases.

\begin{figure}[t]
\centering
\includegraphics[width=\linewidth]{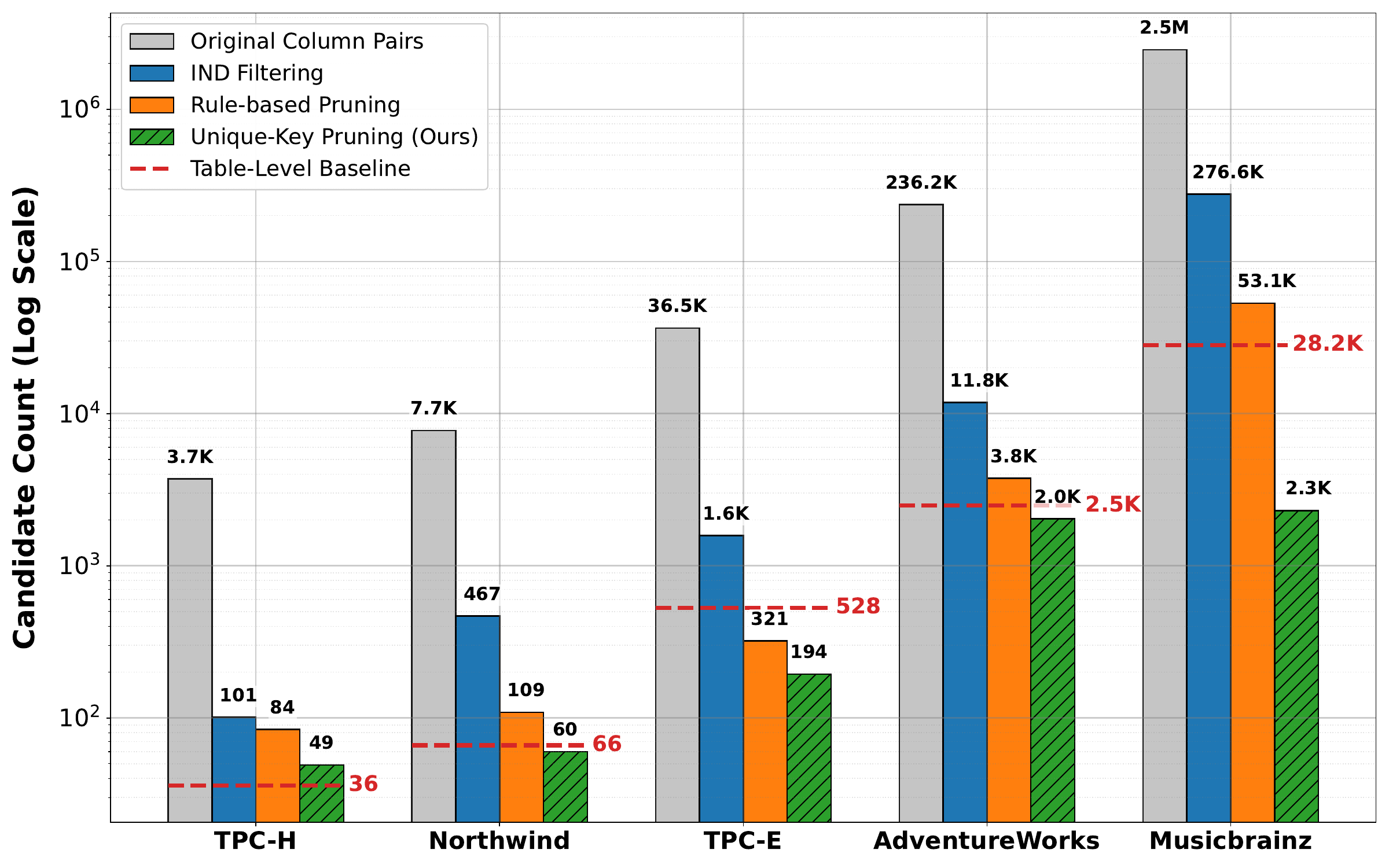}
\caption{Remaining candidate counts after various stages of pruning compared to the table-level baseline.}
\label{fig:pruning_effect}
\end{figure}

\subsection{Analysis of Candidate Pruning}
\label{Analysis of Candidate Pruning Efficiency}
% To demonstrate the effectiveness of our pruning strategy in reducing the FK candidate search space, we analyze the performance from two key perspectives: search space reduction and pruning safety. 
% Figure \ref{fig:pruning_effect} compares the candidate count at each pruning stage against the commonly used table-level baseline, defined as the set of all distinct pairwise table combinations.

We analyze the pruning strategy from two perspectives: search space reduction and pruning safety. Figure~\ref{fig:pruning_effect} shows the number of remaining candidates after each pruning stage, compared to a commonly used table-level baseline defined by all pairwise combinations of tables.

% We analyze the pruning strategy from two complementary perspectives: search space reduction and pruning safety.
% Figure~\ref{fig:pruning_effect} reports the number of remaining candidates after each pruning stage, compared against a commonly used table-level baseline defined by all distinct table pairs.

\paragraph{Search Space Reduction.} 
% Our multi-stage strategy achieves a massive compression of the candidate. 
% The candidate pairs for the large-scale Musicbrainz dataset were reduced from over $2.46 \times 10^6$ to a mere 2,306—a reduction of three orders of magnitude. 
% Notably, our method results in fewer candidates than the commonly used table-level approach in most datasets. This demonstrates that our strategy effectively mitigates the combinatorial explosion, thereby drastically minimizing LLM invocations, ensuring that the subsequent LLM inference remains computationally feasible.
The proposed multi-stage pruning strategy achieves dramatic compression of the FK candidate space.
On the large-scale MusicBrainz dataset, the number of candidate pairs is reduced from over $2.46 \times 10^6$ to only 2,306, corresponding to a reduction of nearly three orders of magnitude.
Across most datasets, the resulting candidate sets are even smaller than those produced by table-level enumeration.
This reduction directly mitigates the combinatorial explosion inherent in large schemas and drastically decreases the number of required LLM invocations, ensuring that downstream reasoning remains computationally feasible.

\paragraph{Pruning Safety.} 
% Crucially, this aggressive filtering does not compromise accuracy. 
% Across all five datasets, the number of pruned true FKs is exactly \textbf{0}. 
% This confirms that our LLM-FK effectively discard the vast majority of irrelevant candidates while retaining 100\% of the ground truth.
% Furthermore, we observe that declared PKs may function as supersets of the selected unique keys (UKs)—typically driven by indexing optimization or domain semantics—while the true FKs reference these specific UK subsets. Consequently, compared to relying on potentially absent or excessively composite declared PKs, our strategy of targeting minimal unique constraints enhances pruning efficacy while strictly preserving true FKs.

Importantly, this aggressive pruning does not sacrifice recall.
Across all five datasets, the number of pruned true FKs is exactly \textbf{zero}, confirming that the strategy preserves 100\% of the ground-truth FKs.
We further observe that declared PKs often act as supersets of the actual referenced unique keys, driven by indexing choices or domain-specific modeling decisions.
In contrast, true FKs frequently reference minimal unique subsets.
By targeting MinUCCs rather than potentially missing or overly composite PKs, LLM-FK achieves more effective pruning while strictly preserving correctness.

\begin{table}[t]
\centering
% 适当增加列间距和行高，防止文字挤在一起看起来太小
\setlength{\tabcolsep}{3.5pt}
\renewcommand{\arraystretch}{1.1}

\resizebox{\columnwidth}{!}{
\begin{tabular}{lccccc}
\toprule
\textbf{Configuration} & \textbf{TPC-H} & \textbf{NWind} & \textbf{TPC-E} & \textbf{AdvWorks} & \textbf{MusicBz} \\
\midrule
\multicolumn{6}{l}{\textit{Analysis A: Refiner Agent}} \\
w/ Injection (Base) & 0.78 & 0.92 & 0.70 & 0.83 & 0.87 \\
w/o Injection& 0.78 & 0.92 & 0.61 & 0.74 & 0.76 \\
\textbf{Perf. Drop ($\Delta$)} & -- & -- & \textbf{-9\%} & \textbf{-9\%} & \textbf{-11\%} \\
\midrule
\multicolumn{6}{l}{\textit{Analysis B: Verifier Agent}} \\
w/ Injection (Full) & 1.00 & 1.00 & 0.93 & 0.94 & 0.95 \\
w/o Injection& 0.78 & 1.00 & 0.89 & 0.94 & 0.95 \\
\textbf{Perf. Drop ($\Delta$)} & \textbf{-22\%} & -- & \textbf{-4\%} & -- & -- \\
\bottomrule
\end{tabular}
}
\caption{Ablation study on the impact of Self-Augmented Domain Knowledge Injection.}
\label{tab:ablation_knowledge}
\end{table}

\subsection{Analysis of Domain Knowledge Injection}
\label{sec:impact_domain_knowledge}
% To evaluate the contribution of Self-Augmented Domain Knowledge Injection, we selectively remove it from the Refiner and the Verifier. Table \ref{tab:ablation_knowledge} summarizes the F1-score performance impact.

% Experimental results confirm its efficacy across both agents. For Refiner, the absence of injected knowledge significantly impairs performance on complex databases like AdventureWorks and Musicbrainz (dropping by $9\%$ and $11\%$, respectively). This decline arises because isolated column pairs in large-scale schemas lack sufficient context, leading LLMs to misinterpret domain semantics without the guidance of injected knowledge. Conversely, For Verifier, this injection is pivotal for smaller datasets like TPC-H, where its removal leads to a significant $22\%$ drop in F1-score. In such compact schemas, global semantic information facilitates a direct comparison among conflicting candidates to effectively distinguish true FKs from structurally similar but incorrect ones. Notably, for TPC-E, whose schema is characterized by widespread English abbreviations, the injected domain knowledge enhances the semantic understanding of local column pairs to mitigate ambiguity (preventing performance drops of 9\% for Refiner and 4\% for Verifier).
To quantify the contribution of Self-Augmented Domain Knowledge Injection, we selectively disable it in the Refiner and the Verifier.
Table~\ref{tab:ablation_knowledge} reports the resulting F1-score changes of each agent's output.

The results confirm that domain knowledge injection is beneficial across both agents.
For the Refiner, removing injected knowledge leads to notable performance drops on AdventureWorks and MusicBrainz (9\% and 11\%, respectively).
This degradation arises because isolated column pairs lack sufficient semantic context in large schemas, causing LLMs to misinterpret entity relationships.
For the Verifier, domain knowledge is especially critical on smaller datasets such as TPC-H, where its removal results in a 22\% F1-score drop.
In compact schemas, global semantic cues enable explicit differentiation among competing candidates during conflict resolution.
On TPC-E, which is dominated by abbreviated identifiers, injected domain knowledge consistently alleviates semantic ambiguity, preventing performance degradation for both agents.
These findings highlight the complementary role of global semantic grounding in both local inference and global consistency enforcement.

\begin{table}[t]
\centering
% 适当增加列间距，因为去掉了竖线，现在有空间了
\setlength{\tabcolsep}{3.5pt} 
\renewcommand{\arraystretch}{1.1} 

\resizebox{\columnwidth}{!}{
\begin{tabular}{llccccc}
\toprule
\textbf{Scenario} & \textbf{Model} & \textbf{TPC-H} & \textbf{NWind} & \textbf{TPC-E} & \textbf{AdvWorks} & \textbf{MusicBz} \\
\midrule
% --- Scenario 1 ---
\multirow{3}{*}{\shortstack[l]{\textbf{Missing}\\\textbf{Data}}} 
& w/o MP-CoT & 0.89 & \textbf{1.00} & 0.88 & 0.90 & 0.93 \\
& \textbf{Full MP-CoT} & \textbf{1.00} & \textbf{1.00} & \textbf{0.92} & \textbf{0.91} & \textbf{0.96} \\
\cmidrule(l){2-7} 
& \textit{Improv. ($\Delta$)} & \textit{+11\%} & \textit{--} & \textit{+4\%} & \textit{+1\%} & \textit{+3\%} \\
\midrule

% --- Scenario 2 ---
\multirow{3}{*}{\shortstack[l]{\textbf{Obscure}\\\textbf{Semantics}}} 
& w/o MP-CoT & 0.74 & \textbf{1.00} & 0.90 & 0.82 & 0.86 \\
& \textbf{Full MP-CoT} & \textbf{0.78} & \textbf{1.00} & \textbf{0.92} & \textbf{0.83} & \textbf{0.89} \\
\cmidrule(l){2-7} 
& \textit{Improv. ($\Delta$)} & \textit{+4\%} & \textit{--} & \textit{+2\%} & \textit{+1\%} & \textit{+3\%} \\
\midrule

% --- Scenario 3 ---
\multirow{3}{*}{\shortstack[l]{\textbf{Anonymous}\\\textbf{Schema}}} 
& w/o MP-CoT & 0.31 & 0.83 & 0.56 & 0.28 & 0.38 \\
& \textbf{Full MP-CoT} & \textbf{0.67} & \textbf{0.92} & \textbf{0.71} & \textbf{0.53} & \textbf{0.76} \\
\cmidrule(l){2-7} 
& \textit{Improv. ($\Delta$)} & \textit{+36\%} & \textit{+9\%} & \textit{+15\%} & \textit{+25\%} & \textit{+38\%} \\
\bottomrule
\end{tabular}
}
\caption{Robustness analysis of Multi-Perspective CoT Reasoning across different challenging scenarios.}
\label{tab:ablation_mpcot}
\end{table}

\subsection{Analysis of Multi-Perspective CoT}
We evaluate the robustness of \textbf{Multi-Perspective CoT (MP-CoT) Reasoning} under three challenging real-world scenarios:
(1) \textbf{Missing Data} (empty tables),
(2) \textbf{Obscure Semantics} (Chinese pinyin abbreviations), and
(3) \textbf{Anonymous Schema} (schema names unavailable).
These scenarios simulate irregular conditions commonly encountered in partial or degraded database schemas.
We compare the F1-scores of the full MP-CoT framework against a variant without MP-CoT in Table~\ref{tab:ablation_mpcot}.

The results yield several insights.
\textbf{First}, removing MP-CoT consistently degrades performance across all datasets, validating its necessity.
The effect is most pronounced under Anonymous Schema conditions, where MP-CoT improves F1-score by up to 38\% on MusicBrainz.
\textbf{Second}, MP-CoT provides strong robustness by allowing remaining perspectives to compensate when one signal is unavailable.
Even in the presence of missing data, where heuristic-based methods typically fail, F1-scores remain above 90\% across all datasets.
% \textbf{Finally}, the analysis confirms that these reasoning perspectives are complementary and non-redundant. Each perspective—syntactic, semantic, and data-driven—compensates for the others' limitations, ensuring resilience against diverse forms of schema degradation.
% the analysis confirms that these reasoning perspectives are complementary and non-redundant.
% Each perspective compensates for specific deficits: syntactic and semantic cues mitigate missing data, while data-driven reasoning becomes pivotal when schema semantics are obscure or names are anonymous. 
% Specifically, syntax and semantics compensate for missing data, whereas syntax and data reasoning mitigate obscure semantics, and data reasoning becomes pivotal when schema names are absent. 
\textbf{Finally}, the analysis further confirms that these reasoning perspectives are complementary and non-redundant, ensuring resilience against schema degradation by mutually compensating for limitations:
syntactic and semantic clues address data sparsity, syntactic and statistical evidence mitigates semantic ambiguity, and statistical signals become pivotal when schema names are absent.
% syntax and semantics reasoning address missing data, syntax and data reasoning mitigate obscure semantics, and data reasoning becomes pivotal when schema names are absent.

\begin{table}[t]
\centering
% 1. 增加行高：因为 resizebox 会把所有东西缩小，所以原始行高要大一点（1.15~1.2），
% 这样缩小后看起来才会有呼吸感，不会挤在一起。
\renewcommand{\arraystretch}{1.2}

% 2. 控制列间距：不要太宽，紧凑一点可以让 resizebox 算出来的字号更大。
\setlength{\tabcolsep}{3pt}

\resizebox{\columnwidth}{!}{
% 3. 去掉所有竖线 |，改用 booktabs 风格
\begin{tabular}{l ccc ccc cc} 
\toprule % 顶部粗线
\multirow{2}{*}{\textbf{Dataset}} & \multicolumn{3}{c}{\textbf{Refiner (Pre-CR)}} & \multicolumn{3}{c}{\textbf{Verifier (Post-CR)}} & \multicolumn{2}{c}{\textbf{Gains}} \\
% \cmidrule(lr){...} 是断开的横线，左右留白(lr)，视觉上非常清晰
\cmidrule(lr){2-4} \cmidrule(lr){5-7} \cmidrule(lr){8-9}
 & \textbf{P} & \textbf{R} & \textbf{F} & \textbf{P} & \textbf{R} & \textbf{F} & \textbf{$\Delta$P} & \textbf{$\Delta$F} \\
\midrule % 中间细线
TPC-H    & 0.64 & 1.00 & 0.78 & 1.00 & 1.00 & 1.00 & \textbf{+36\%} & +22\% \\
NWind    & 0.85 & 1.00 & 0.92 & 1.00 & 1.00 & 1.00 & \textbf{+15\%} & +8\% \\
TPC-E    & 0.56 & 0.96 & 0.70 & 0.91 & 0.96 & 0.93 & \textbf{+35\%} & +23\% \\
AdvWorks & 0.71 & 0.99 & 0.83 & 0.93 & 0.96 & 0.94 & \textbf{+22\%} & +11\% \\
MusicBz  & 0.77 & 1.00 & 0.87 & 0.91 & 1.00 & 0.95 & \textbf{+14\%} & +8\% \\
\bottomrule % 底部粗线
\end{tabular}
}
\caption{Ablation study on the impact of Holistic Conflict Resolution.}
\label{tab:ablation_cr}
\end{table}
% \begin{table}[t]
% \centering
% % 1. 进一步减小列间距
% \setlength{\tabcolsep}{3pt} 
% % 2. 将行高设置为 0.95 或 1.0 (1.25 太高了)
% \renewcommand{\arraystretch}{0.95}

% \resizebox{\columnwidth}{!}{
% \begin{tabular}{l|ccc|ccc|cc}
% \hline
% \multirow{2}{*}{\textbf{Dataset}} & \multicolumn{3}{c|}{\textbf{Refiner (Pre-CR)}} & \multicolumn{3}{c|}{\textbf{Final (Post-CR)}} & \multicolumn{2}{c}{\textbf{Gains}} \\
% \cline{2-9}
% & \textbf{P} & \textbf{R} & \textbf{F} & \textbf{P} & \textbf{R} & \textbf{F} & \textbf{$\Delta$P} & \textbf{$\Delta$F} \\
% \hline
% TPC-H      & 0.64 & 1.00 & 0.78 & 1.00 & 1.00 & 1.00 & \textit{\textbf{+36\%}} & \textit{+22\%} \\
% NWind      & 0.85 & 1.00 & 0.92 & 1.00 & 1.00 & 1.00 & \textit{\textbf{+15\%}} & \textit{+8\%} \\
% TPC-E      & 0.56 & 0.96 & 0.70 & 0.91 & 0.96 & 0.93 & \textit{\textbf{+35\%}} & \textit{+23\%} \\
% AdvWorks   & 0.71 & 0.99 & 0.83 & 0.93 & 0.96 & 0.94 & \textit{\textbf{+22\%}} & \textit{+11\%} \\
% MusicBz    & 0.77 & 1.00 & 0.87 & 0.91 & 1.00 & 0.95 & \textit{\textbf{+14\%}} & \textit{+8\%} \\
% \hline
% \end{tabular}
% }
% \caption{Impact of Holistic Conflict Resolution.}
% \label{tab:ablation_cr}
% \end{table}

\subsection{Analysis of Holistic Conflict Resolution}
\label{sec:analysis_cr}

% We evaluate the efficacy of the Holistic Conflict Resolution (CR) by comparing the candidates generated by the Refiner (Pre-CR) with the final output validated by the Verifier (Post-CR). As shown in Table \ref{tab:ablation_cr}, the Refiner is intentionally aggressive to maximize candidate retrieval. While this strategy secures high Recall, it inevitably compromises Precision.
% The Verifier acts as a critical filter for these false positives. It substantially boosts Precision (ranging from +14\% to +36\%) while successfully retaining the high Recall(with correct FKs fully preserved in four databases), effectively ensuring global structural consistency.

We evaluate the effectiveness of the Holistic Conflict Resolution (CR) Strategy by comparing Refiner outputs before conflict resolution (Pre-CR) with the final predictions produced by the Verifier (Post-CR).
As shown in Table~\ref{tab:ablation_cr}, the Refiner is intentionally designed to be aggressive, prioritizing high recall at the cost of lower precision.

The Verifier serves as a critical global filter that resolves structural inconsistencies.
Across all datasets, conflict resolution substantially improves precision, with gains ranging from +14\% to +36\%, while largely preserving recall.
In four out of five datasets, all true FKs identified by the Refiner are retained after CR. 
Notably, all cyclic references encountered are strictly length-2, ensuring that the resolution avoids the cascading complexity associated with global traversals.
These results demonstrate that holistic, schema-level reasoning is essential for reconciling local predictions and ensuring global consistency, enabling LLM-FK to achieve both high accuracy and structural correctness.

\section{Conclusion}
\label{Conclusion}
% We present \textbf{LLM-FK}, an automated multi-agent framework for FK detection in large-scale complex databases. By orchestrating four specialized agents, LLM-FK decomposes the detection task, performs semantically grounded reasoning, and enforces global consistency.
% The Profiler reduces the search space via Unique-Key-Driven Schema Decomposition; the Interpreter and Refiner leverage Self-Augmented Domain Knowledge and Multi-Perspective Chain-of-Thought for robust inference; and the Verifier reconciles local predictions through Holistic Conflict Resolution.
% Extensive experiments show that LLM-FK consistently achieves state-of-the-art performance, outperforming heuristic, learning-based, and generic LLM baselines. By explicitly modeling semantic reasoning, structural constraints, and global consistency, LLM-FK provides an accurate, efficient, and robust solution for automated FK discovery, offering a practical foundation for downstream database management and analytics tasks.

We present \textbf{LLM-FK}, the first multi-agent framework for FK detection in large-scale complex databases. By orchestrating four specialized agents, LLM-FK decomposes the detection task, performs semantically grounded reasoning, and enforces global consistency. Specifically, the Profiler reduces the search space via Unique-Key-Driven Schema Decomposition Strategy; the Interpreter and Refiner leverage Self-Augmented Domain Knowledge Injection and Multi-Perspective CoT Reasoning for deep and robust inference; and the Verifier reconciles local predictions through Holistic Conflict Resolution Strategy. Extensive experiments show that LLM-FK consistently achieves SOTA performance, providing an accurate, efficient, and robust solution for FK discovery that serves as a practical foundation for downstream database management and analytics tasks.

\section*{Limitations}
\label{LIMITATIONS}
While our experiments cover five widely adopted benchmarks spanning commerce, retail, finance, manufacturing, and entertainment, and demonstrate LLM-FK's strong generalization and robustness, there remain unexplored database domains with extremely large-scale or highly heterogeneous schemas. 
Importantly, LLM-FK is designed to be domain-agnostic and relies solely on schema-inherent signals rather than domain-specific knowledge, ensuring that its performance does not depend on particular conventions or languages. 
Moreover, its modular multi-agent design allows seamless extension to new databases without additional training or supervision. 
Combined with our candidate pruning and structured reasoning strategies, LLM-FK maintains computational feasibility and robust generalization even in complex schemas, making it broadly applicable while leaving room for future evaluations on specialized domains.

\section*{Acknowledgments}
\label{Acknowledgments}
This work was supported by the National Key Laboratory of Data Space Technology and System.

\bibliography{main}
% \newpage
\appendix
\label{sec:appendix}
\section{More Related Work}
\subsection{Metadata Discovery}
\label{Metadata Discovery}
% General Context and Complexity
% Many research has employed various search strategies to optimize the discovery of UCCs and INDs, rendering the extraction of these metadata computationally feasible in practice\citep{abedjan2015profiling, de2002efficient, dursch2019inclusion}. Representative algorithms include Spider \citep{bauckmann2006efficiently}, which targets single-column INDs using a disk-based sort-merge join, and Binder \citep{papenbrock2015divide}, which extends detection to multi-column INDs via a hash-based divide-and-conquer strategy.
% Furthermore, to address data quality issues such as dirty data in real-world scenarios, several approaches have been proposed to discover approximate INDs\citep{tschirschnitz2017detecting, kaminsky2023discovering}.
% The metadata discovery module of LLM-FK can be instantiated by these established algorithms, allowing the framework to be tailored to specific objectives, such as prioritizing computational efficiency or maximizing tolerance to data anomalies.
Extensive research has investigated efficient strategies for discovering Unique Column Combinations (UCCs) and Inclusion Dependencies (INDs), making the extraction of such metadata computationally feasible even for large databases~\citep{de2002efficient, abedjan2015profiling, dursch2019inclusion}. Representative algorithms include \textbf{Spider}~\citep{bauckmann2006efficiently}, which efficiently identifies single-column INDs using a disk-based sort-merge join, and \textbf{Binder}~\citep{papenbrock2015divide}, which generalizes the approach to multi-column INDs via a hash-based divide-and-conquer strategy. 

To address practical challenges such as dirty or incomplete data, approximate IND discovery techniques have also been proposed~\citep{tschirschnitz2017detecting, kaminsky2023discovering}. 
The metadata discovery process within LLM-FK is fully compatible with these established algorithms, allowing practitioners to tailor the framework according to specific objectives—such as prioritizing computational efficiency or maximizing robustness to data anomalies—without modifying the overall reasoning pipeline.

% Recently, multi-agent systems have demonstrated promising potential in tackling complex reasoning tasks by facilitating collaborative problem-solving among specialized agents \citep{guo2024large, zhang2024sequential, guan2024advancing}. While existing research has explored multi-agent frameworks in the tabular domain, these methods predominately focus on table reasoning or question answering, mainly targeting the specific content of single table or partial tables\citep{wang2025mac, yu2025table, sui2025chain}. This scope differs significantly from the challenge of FK detection, which requires a holistic structural analysis at the entire database level. To the best of our knowledge, our proposed LLM-FK represents the first attempt to introduce an LLM-based multi-agent framework specifically for FK detection, offering a perspective on addressing the challenges inherent in detecting constraints within large and complex database schemas.

\subsection{Multi-Agent Systems}

Multi-agent systems have recently shown strong potential for solving complex reasoning tasks by enabling collaboration among specialized agents~\citep{guan2024advancing, guo2024large, zhang2024sequential}. 
% In the tabular domain, existing multi-agent frameworks primarily focus on table-level reasoning or question answering, often limited to single tables or partial schemas~\citep{zhou2025efficient, wang2025mac, yu2025table}.
In the tabular domain, existing multi-agent frameworks primarily target downstream applications like Text-to-SQL and table question answering~\citep{xie2024mag, sui2025chain, wang2025mac, yu2025table, zhou2025efficient}. These tasks generally assume a known schema and rely on content matching and local schema linking to satisfy specific queries. 

In contrast, FK detection presents a fundamentally different challenge: it requires holistic structural reasoning across an entire database schema. To the best of our knowledge, LLM-FK is the first framework to leverage multi-agent collaboration specifically for FK detection, demonstrating how specialized agents can jointly address the combinatorial and inconsistency challenges inherent in large, complex databases. This design illustrates the potential of multi-agent reasoning beyond content-oriented tasks, extending it to structural inference and schema-level constraint discovery.

\subsection{FK Detection in Modern Data Ecosystems}

While FKs are fundamental to relational database theory, their explicit definition is frequently absent in practice, making FK detection a critical task for downstream data operations. The urgency and practical significance of this problem in contemporary settings can be summarized across two key dimensions:

\paragraph{Ubiquity of Missing FKs.} In modern data environments, the absence of explicit FKs has become a widespread norm driven by evolving engineering practices and architectural shifts. To avoid performance degradation during high-concurrency operations, developers frequently omit physical FKs, shifting constraint management entirely to the application layer~\citep{AlibabaGitbookAJCG}, making these implicit constraints highly prone to loss during database migration or analytical extraction~\citep{jiang2020holistic}. Furthermore, distributed architectures—such as the ``database-per-service'' pattern in microservices~\citep{newman2021building} and schema-on-read data lakehouses~\citep{armbrust2021lakehouse}—inherently restrict strict, cross-database FK constraints. Empirical studies corroborate this scarcity, revealing that a significant majority of real-world open-source web applications and relational databases completely lack explicit FK definitions~\citep{agarwal2021retrofitting, christopher2021schemadb, dohmen2024schemapile}.

\paragraph{Critical Role in Advanced Data Applications.} Beyond classic data management tasks like data integration and schema reverse engineering, FK detection is increasingly pivotal in contemporary data ecosystems. In complex, heterogeneous environments, identifying FK relationships is highly valuable for streamlining data discovery and bolstering governance efficiency~\citep{fernandez2018aurum, hai2023data}. Moreover, accurately inferring these structural dependencies provides an essential foundation for empowering automated Business Intelligence (BI) modeling~\citep{lin2023auto}, as FK detection closely aligns with the foundational steps of BI model construction and enables automatic generation of schema graphs for downstream analytical pipelines. In the current era of LLM-driven database interfaces, explicit FKs provide deterministic join paths that have been empirically proven to substantially enhance the accuracy of Text-to-SQL parsing~\citep{gao2023text}. More broadly, multi-table Question Answering (QA) has recently emerged as a focal point in LLM-driven table understanding\citep{zhu2024autotqa, wang2025plugging}, where FK detection is explicitly identified as a prerequisite for resolving multi-hop complex questions across diverse tables~\citep{wu2025mmqa}. This positions robust FK detection as a key enabler for seamlessly integrating structured relational semantics into emerging LLM-based table reasoning systems.

\section{More Implementation Details}
\label{More Implementation Details}
% 这部分主要就是把文章写的部分全部拿过来
\subsection{Detailed Elaboration of Multi-Agent Collaboration}
\label{sec:agent_collaboration}
% To automate the reasoning required for FK detection, we orchestrate four specialized agents—Profiler, Interpreter, Refiner, and Verifier—to collaboratively process the database. We employ specific instructions $\mathcal{I}$ to prompt the LLM, denoted as $\pi$, to execute distinct cognitive operations. 
% Formally, we define the role and operational formulation of each agent as follows:
To automate the reasoning required for FK detection, LLM-FK orchestrates four specialized agents—\textbf{Profiler}, \textbf{Interpreter}, \textbf{Refiner}, and \textbf{Verifier}—to collaboratively execute a database-wide FK detection task. Each agent executes distinct cognitive operations via specific LLM instructions $\mathcal{I}$, denoted as $\pi$. Formally, the roles and operational formulations are as follows:

\paragraph{Profiler ($\mathcal{A}_{p}$).} 
% The Profiler serves as the entry point of the entire agentic workflow, implementing a Unique-Key-Driven Schema Decomposition to address schema decomposition and search space explosion. Given the database $\mathcal{D}$, the set of identified INDs denoted as $\mathcal{J}$ (which are pre-filtered to satisfy data type rule constraints), and the set of MinUCCs $\mathcal{U}$ derived specifically from the referenced tables within $\mathcal{J}$ via metadata profiling tools, the Profiler iterates through each referenced table, denoted as $T$. For each $T$, utilizing its specific subset of MinUCCs $\mathcal{U}_T \subseteq \mathcal{U}$, the agent identifies the most suitable referenced unique key $k$ guided by the instruction  $\mathcal{I}_{\mathcal{A}_{p}}$. Formally, the key selection process for a target table is defined as:
The Profiler serves as the entry point of the agent workflow, implementing a \textbf{Unique-Key-Driven Schema Decomposition} to mitigate search space explosion. Given a database $\mathcal{D}$, the set of identified INDs $\mathcal{J}$ (pre-filtered by data type rules), and the set of MinUCCs $\mathcal{U}$ derived from referenced tables in $\mathcal{J}$, the Profiler iterates over each referenced table $T$. Using its subset of MinUCCs $\mathcal{U}_T \subseteq \mathcal{U}$, it selects the most suitable referenced unique key $k$ guided by $\mathcal{I}_{\mathcal{A}_{p}}$:
\begin{equation}
    k = \pi(T, \mathcal{U}_T, \mathcal{I}_{\mathcal{A}_{p}}),
\end{equation}
% Subsequently, the Profiler aggregates these selected unique keys from all referenced tables into a unique key set $\mathcal{K}$. Based on this set, the search space is deterministically pruned. The set of structurally valid candidate FK pairs, denoted as $\Psi_{cand} = \{(c_f, c_p) \in \mathcal{J} \mid c_p \in \mathcal{K}\}$, is derived by filtering the set $\mathcal{J}$ against the selected unique keys within $\mathcal{K}$.
These keys are aggregated into a set $\mathcal{K}$, which deterministically prunes the search space. The structurally valid candidate FK pairs are then:

\begin{equation}
    \Psi_{cand} = \{(c_f, c_p) \in \mathcal{J} \mid c_p \in \mathcal{K}\},
\end{equation}

\paragraph{Interpreter ($\mathcal{A}_{i}$).} 
% The Interpreter performs the Self-Augmented Domain Knowledge Injection. To mitigate the ambiguity of abstract schema symbols, it analyzes the pruned candidate set $\Psi_{cand}$ to distill global semantic context. Utilizing the set of unique table names $\mathcal{N}_{\Psi_{cand}}$ derived from the column pairs in $\Psi_{cand}$ and the instruction $\mathcal{I}_{\mathcal{A}_{i}}$, the Interpreter operates as:
% \begin{equation}
%     \mathcal{K}_{dom} = \pi(\mathcal{N}_{\Psi_{cand}}, \mathcal{I}_{\mathcal{A}_{i}}).
% \end{equation}
% The resulting $\mathcal{K}_{dom}$ encapsulates the deduced domain knowledge (e.g., application topic, core entity concepts). This global context is computed once and shared downstream to ensure semantic consistency across local reasoning tasks.
The Interpreter performs \textbf{Self-Augmented Domain Knowledge Injection} to reduce schema ambiguities. Analyzing $\Psi_{cand}$, it derives global domain knowledge $\mathcal{K}_{dom}$ from the set of table names $\mathcal{N}_{\Psi_{cand}}$:

\begin{equation}
    \mathcal{K}_{dom} = \pi(\mathcal{N}_{\Psi_{cand}}, \mathcal{I}_{\mathcal{A}_{i}}),
\end{equation}

which is shared downstream to ensure semantic consistency across local reasoning tasks.

\paragraph{Refiner ($\mathcal{A}_{r}$).} 
The Refiner validates each candidate pair $\psi = (c_f, c_p) \in \Psi_{cand}$ using \textbf{Multi-Perspective CoT Reasoning}, integrating domain knowledge $\mathcal{K}_{dom}$ and serialized structural representations $S_\psi$ derived from SQL queries:

\begin{equation}
    p_{\psi} = \pi(\psi, S_\psi, \mathcal{K}_{dom}, \mathcal{I}_{\mathcal{A}_{r}}),
\end{equation}

where $p_{\psi} \in \{0, 1\}$ indicates whether $\psi$ constitutes a true FK. We denote the set of positively validated candidates as:

\begin{equation}
    \Psi_{pos} = \{\psi \in \Psi_{cand} \mid p_{\psi} = 1\},
\end{equation}

\paragraph{Verifier ($\mathcal{A}_{v}$).} 
The Verifier ensures global coherence via the \textbf{Holistic Conflict Resolution Strategy}. Initialized with $\Psi_{pos}$, it identifies structural violations (e.g., multiple or cyclic references) within the aggregated schema topology. For any identified conflict set $\Psi_{conflict}$, it determines the valid subset $\Psi_{resolved}$ by leveraging domain knowledge $\mathcal{K}_{dom}$ and structural representations $S_{\Psi_{conflict}}$:

\begin{equation}
    \Psi_{resolved} = \pi(\Psi_{conflict}, S_{\Psi_{conflict}}, \mathcal{K}_{dom}, \mathcal{I}_{\mathcal{A}_{v}}).
\end{equation}

% This resolution mechanism refines the global schema topology by eliminating invalid candidate FKs, ensuring that the schema converges to a conflict-free state to yield the final consistent FK set $\Phi$.
This resolution mechanism refines the global schema topology by eliminating invalid candidate FKs, ensuring convergence to a conflict-free state and yielding the final consistent FK set $\Phi$.
% \paragraph{Verifier ($\mathcal{A}_{v}$).} 
% The Verifier ensures global coherence via the \textbf{Holistic Conflict Resolution Strategy}. Initialized with $\Psi_{pos}$, it iteratively detects structural violations, specifically targeting conflicts like multiple references and cyclic references. In each step, for a targeted conflict set $\Psi_{conflict}$, it resolves the contention to derive the valid subset $\Psi_{resolved}$ (i.e., retaining the most plausible connections while breaking loops) by leveraging domain knowledge $\mathcal{K}_{dom}$ and structural representations $S_{\Psi_{conflict}}$:

% \begin{equation}
%     \Psi_{resolved} = \pi(\Psi_{conflict}, S_{\Psi_{conflict}}, \mathcal{K}_{dom}, \mathcal{I}_{\mathcal{A}_{v}}).
% \end{equation}

% This resolution progressively updates the global schema topology, repeating until the graph converges to an acyclic and unambiguous state, thereby establishing the final consistent FK set $\Phi$.

\paragraph{Multi-Agent Collaboration.} 
% The comprehensive workflow of the multi-agent collaboration is formally defined in Algorithm~\ref{alg:llm_fk_pipeline}. Furthermore, the specific prompts corresponding to the core strategies employed by each agent are detailed in Appendix~\ref{Prompts of LLM-FK and Case Study}.
The overall workflow is formally defined in Algorithm~\ref{alg:llm_fk_pipeline}. Detailed prompts for each agent are provided in Appendix~\ref{Prompts of LLM-FK and Case Study}.

\subsection{Detailed Elaboration of Profiler}
\subsubsection{IND-based and Rule-based Pruning}
To effectively reduce the search space, the Profiler enforces two fundamental rules for pruning:

% \begin{itemize}
%     \item \textbf{Type Compatibility:} Candidate columns must strictly share the same data type.
%     \item \textbf{Exclusion of Unsuitable Types:} Columns with structurally inappropriate types for FKs, such as floating-point numbers and Booleans, are automatically filtered out.
% \end{itemize}
\begin{itemize}
    \item \textbf{Type Compatibility:} Candidate columns must strictly share the same data type.
    \item \textbf{Exclusion of Unsuitable Types:} Columns representing \textit{continuous measurements} (e.g., \texttt{float}, \texttt{decimal}) or \textit{binary states} (e.g., \texttt{bool}, \texttt{bit}) are automatically filtered out as they semantically lack identifying properties.
\end{itemize}

Beyond these foundational constraints, our IND-based and rule-based mechanisms are designed with significant flexibility and extensibility. The IND-based validation supports adaptive algorithm selection (see Appendix \ref{Metadata Discovery}) to optimize the balance between execution efficiency and robustness. Furthermore, additional rules---such as filtering empty tables or columns and applying specific cardinality thresholds---can be seamlessly integrated into the rule-based pruning.

\subsubsection{Unique Key Selection}

Optimal unique key selection leverages LLM prompts guided by formalized judgment criteria (see Table~\ref{tab:uk_selection_rules}), including uniqueness validity, column ordinality, naming conventions, type suitability, value conciseness, minimal composition, and semantic alignment. 
By grounding the selection process in these formalized rules, the LLM can effectively discern the most appropriate referenced unique key from the set of MinUCCs. This approach leverages the model's reasoning capabilities to prioritize candidates that not only satisfy structural constraints but also align with the logical conventions of the database schema, ensuring the selected keys are semantically meaningful.

\begin{table*}[t]
\centering
% 1. 增加行间距，默认是1，改为1.2或1.3会让表格不那么挤
\renewcommand{\arraystretch}{1.3} 
% 2. 使用 tabularx 替代 resizebox
% l 代表第一列根据内容自适应宽度
% X 代表第二列自动填充剩余空间并换行
\begin{tabularx}{\textwidth}{l X} 
\toprule
\textbf{Criterion} & \textbf{Description} \\
\midrule
\textbf{Uniqueness Validity} & The candidate must uniquely identify a specific tuple within the table. \\
\textbf{Column Ordinality} & Columns positioned earlier in the table definition are prioritized. \\
\textbf{Naming Convention} & Preference is given to column names containing typical identifiers (e.g., ``id'', ``key'', ``code''). \\
\textbf{Type Suitability} & The data type is typically an integer or a string, adhering to standard key definitions. \\
\textbf{Value Conciseness} & The data values should be concise and human-readable (short text length). \\
\textbf{Minimal Composition} & Candidates with fewer constituent columns (atomic keys) are preferred over complex composite keys. \\
\textbf{Semantic Alignment} & The key must logically represent the entity according to the semantic context of the table. \\
\bottomrule
\end{tabularx}
\caption{Judgment criteria for unique key selection.}
\label{tab:uk_selection_rules}
\end{table*}

% \begin{table*}[t]
% \centering
% \resizebox{\textwidth}{!}{%
% \begin{tabular}{ll}
% \toprule
% \textbf{Criterion} & \textbf{Description} \\
% \midrule
% \textbf{Uniqueness Validity} & The candidate must uniquely identify a specific tuple within the table. \\
% \textbf{Column Ordinality} & Columns positioned earlier in the table definition are prioritized. \\
% \textbf{Naming Convention} & Preference is given to column names containing typical identifiers (e.g., ``id'', ``key'', ``code''). \\
% \textbf{Type Suitability} & The data type is typically an integer or a string, adhering to standard key definitions. \\
% \textbf{Value Conciseness} & The data values should be concise and human-readable (short text length). \\
% \textbf{Minimal Composition} & Candidates with fewer constituent columns (atomic keys) are preferred over complex composite keys. \\
% \textbf{Semantic Alignment} & The key must logically represent the entity according to the semantic context of the table. \\
% \bottomrule
% \end{tabular}%
% }
% \caption{Judgment Criteria for Unique Key Selection.}
% \label{tab:uk_selection_rules}
% \end{table*}

\subsection{Detailed Elaboration of Interpreter}
% This section elaborates on the Self-Augmented Domain Knowledge Injection mechanism employed by the Interpreter. As illustrated in Figure~\ref{fig:interpreter_workflow}, this approach adopts a summarize-then-reason paradigm, where the LLM first synthesizes a global domain profile from collective table semantics to prime itself with a unified context. This strategy enhances internal consistency by establishing a shared semantic foundation before addressing specific local inferences. Crucially, the framework achieves self-augmentation by injecting these self-generated domain insights back into the reasoning context, thereby enriching the semantic representation of abstract schema symbols with concrete domain knowledge. Notably, the entire process achieves full automation, autonomously deriving macroscopic domain insights without the need for manual annotations or external prompts.
% The Interpreter implements a summarize-then-reason strategy (see Figure~\ref{fig:interpreter_workflow}). LLMs synthesize a global domain profile from table semantics, then inject these self-generated insights back into local reasoning tasks. This ensures semantic consistency and fully automated domain knowledge enrichment without external manual annotations.

\begin{table}[t]
    \centering
    \small
    % 将总宽度设为 \linewidth，第一列左对齐(l)，第二列自动适应并换行(X)
    \begin{tabularx}{\linewidth}{l X} 
        \toprule
        \textbf{Type} & \textbf{Information} \\
        \midrule
        \textbf{Schema Definitions} & Table name \\
         & Column name \\
        \midrule
        \textbf{Statistical Metadata} & Ordinal position \\
         & Data type \\
         & Average value text length \\
         & Number of distinct values \\
         & Number of rows in table \\
         & Cardinality ratio (distinct values / table rows) \\
         & Minimum value \\
         & Maximum value \\
        \midrule
        \textbf{Inter-column Dependencies} & Coverage ratio \\
         & Table size ratio \\
         & Out-of-range ratio \\
        \midrule
        \textbf{Sample Data} & Example data from table (first 5 rows) \\
        \bottomrule
    \end{tabularx}
    \caption{Serialized structural representations for Multi-Perspective CoT Reasoning.}
    \label{tab:sampling_augmentation_info}
\end{table}

\begin{figure}
  \includegraphics[width=\columnwidth]{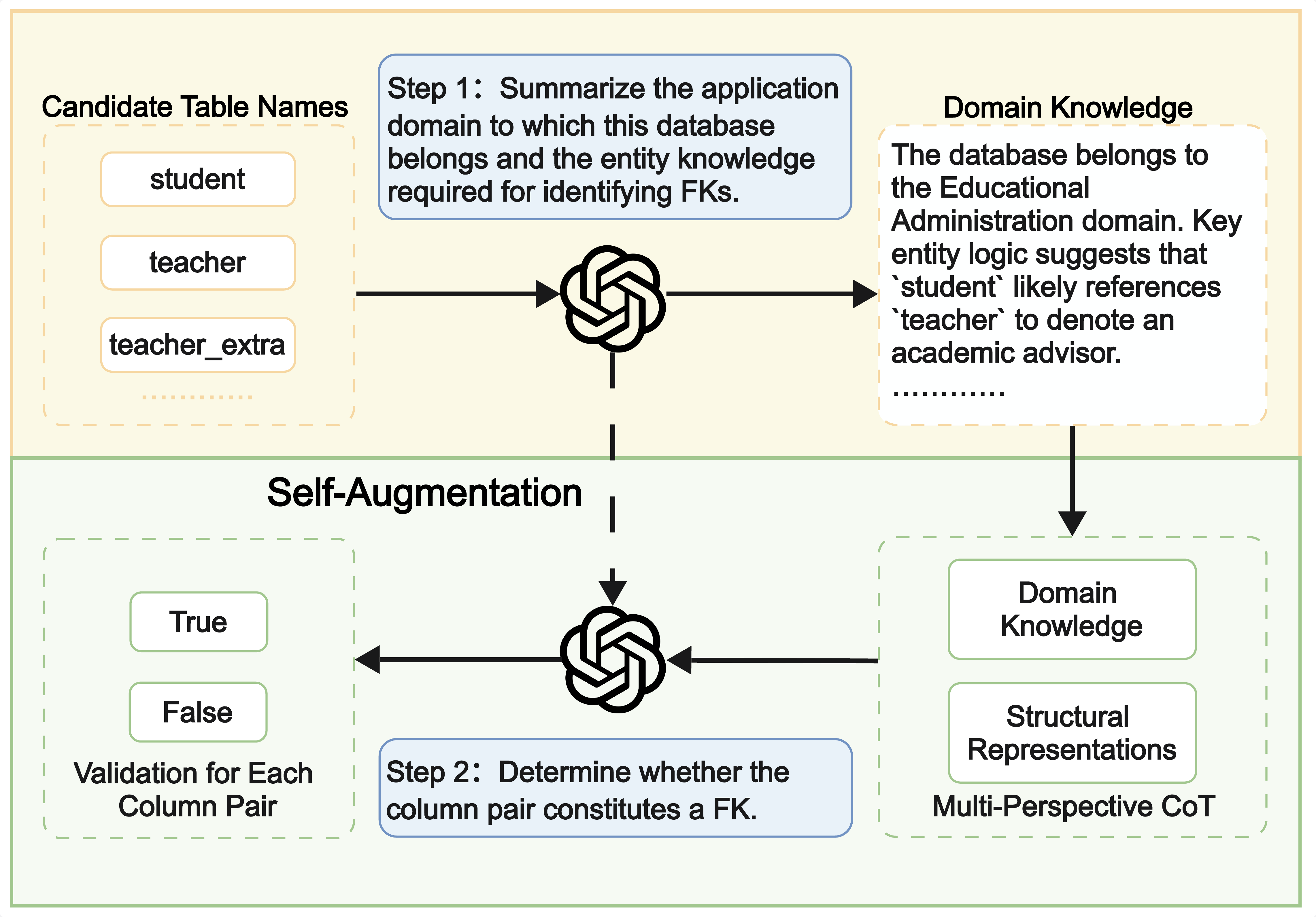}
\caption{Illustration of the Self-Augmented Domain Knowledge Injection.}
\label{fig:interpreter_workflow}       
\end{figure}

The Interpreter implements a summarize-then-reason strategy (see Figure~\ref{fig:interpreter_workflow}), which mirrors the human cognitive process of acquiring a holistic understanding of the database domain before making specific judgments. In this workflow, the LLM first synthesizes a global domain profile from collective table semantics to prime itself with a unified context. Crucially, these self-generated insights are injected back into local reasoning tasks, allowing the framework to self-augment abstract schema symbols with concrete domain knowledge. This global-to-local paradigm ensures semantic consistency by grounding isolated inferences in a shared foundation, while simultaneously achieving full automation—autonomously deriving macroscopic insights without reliance on external manual annotations.

\subsection{Detailed Elaboration of Refiner}
\paragraph{Structural Representations Extraction.}
% To support the subsequent deep inference, the Refiner first constructs a information basis by transforming raw table data into serialized, high-density structural representations. Utilizing SQL queries, the framework extracts signals while filtering out noise. As detailed in Table~\ref{tab:sampling_augmentation_info}, this structural serialization serves as the informational basis for the reasoning, ensuring that the subsequent multi-perspective analysis is grounded in precise schema definitions, statistical facts, and concrete data samples.
The Refiner constructs serialized structural representations by formulating SQL queries on raw table data (see Table~\ref{tab:sampling_augmentation_info}), including schema definitions, statistical metadata, inter-column dependencies, and sample data. This ensures that multi-perspective reasoning is grounded in precise schema and data evidence.

% \begin{table*}[htbp] % 【关键修改】使用 table* 让表格跨越双栏
%     \centering
%     \small
%     % 【关键修改】宽度改为 \textwidth 以填满整个页面宽度
%     \begin{tabularx}{\textwidth}{l X} 
%         \toprule
%         \textbf{Type} & \textbf{Information} \\
%         \midrule
%         \textbf{Schema Definitions} & Table name \\
%          & Column name \\
%         \midrule
%         \textbf{Statistical Metadata} & Column position in table \\
%          & Data type \\
%          & Average value text length \\
%          & Number of distinct values \\
%          & Number of rows in table \\
%          & Cardinality ratio (distinct values/table rows) \\
%          & Minimum value \\
%          & Maximum value \\
%         \midrule
%         \textbf{Inter-column Dependencies} & Coverage ratio \\
%          & Table size ratio \\
%          & Out-of-range ratio \\
%         \midrule
%         \textbf{Sampled Data} & Example data from table (first 5 rows) \\
%         \bottomrule
%     \end{tabularx}
%     \caption{Common Sampling Data and Augmented Information for FK Reconstruction.}
%     \label{tab:sampling_augmentation_info}
% \end{table*}

\paragraph{Multi-Perspective CoT Reasoning.}
% Building upon these structured inputs, the Refiner orchestrates a Multi-Perspective CoT framework to realize structural-semantic synergy. As illustrated in Figure~\ref{fig:refiner_cot}, this mechanism simulates human expert reasoning by integrating three complementary perspectives to validate candidate pairs. The inference process is structured as follows:

\begin{figure}[t]
\centering
\includegraphics[width=\linewidth]{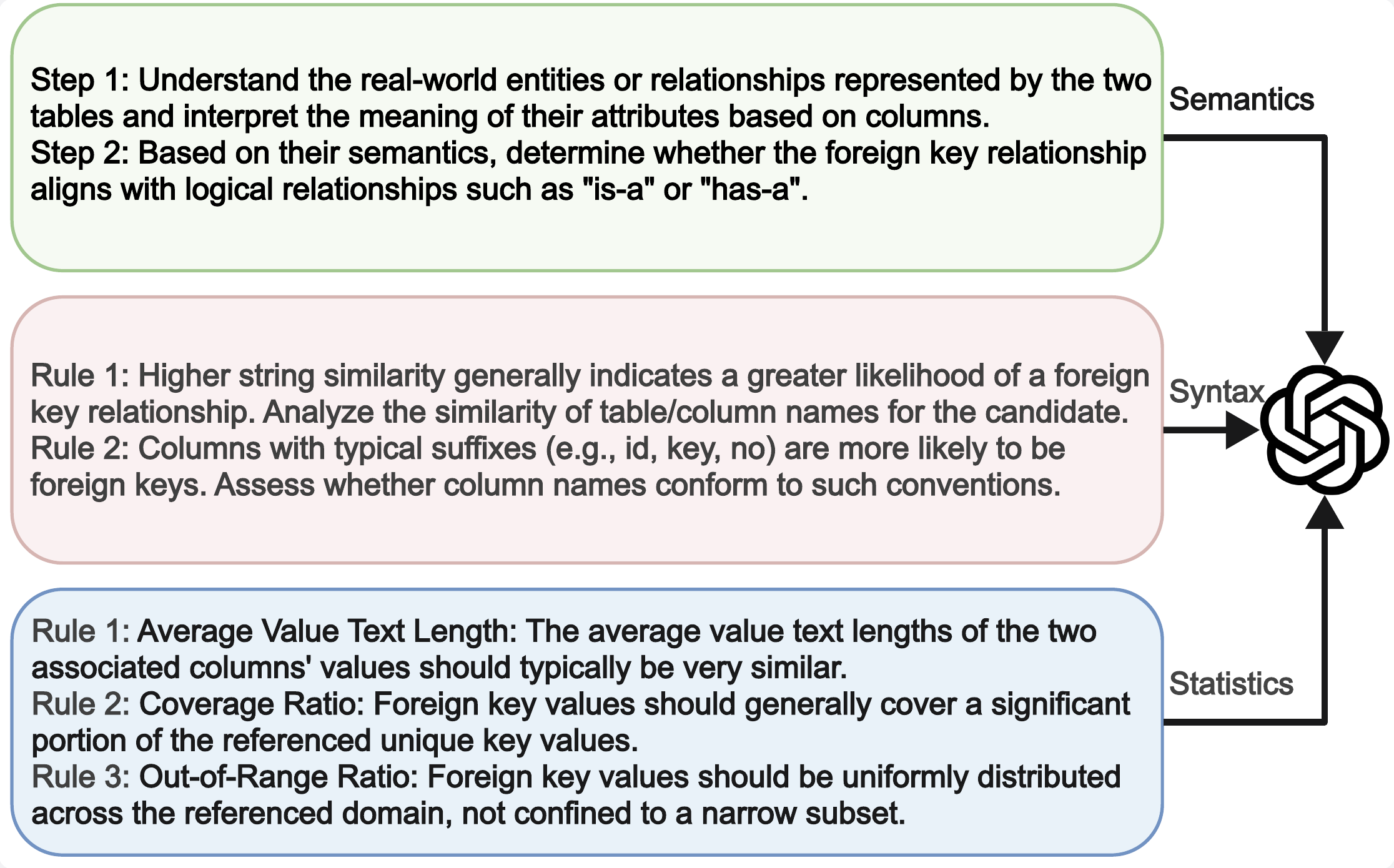}
\caption{Illustration of the Multi-Perspective CoT Reasoning.}
\label{fig:refiner_cot}
\end{figure}

% \begin{itemize}
%     \item \textbf{Semantic Perspective:} Deduces real-world entity relationships by interpreting schema concepts within the macroscopic domain context, effectively bridging abstract symbols with application-specific logic.
%     \item \textbf{Syntactic Perspective:} Evaluates naming conventions and string similarities across table and column names to identify potential linguistic matches.
%     \item \textbf{Data Perspective:} Analyzes statistical distributions and content overlaps---derived directly from statistical metadata and inter-column dependencies---to detect structural compatibility and inclusion relationships.
% \end{itemize}
The Refiner integrates three reasoning perspectives (see Figure~\ref{fig:refiner_cot}):

\begin{itemize}
    \item \textbf{Semantic:} Maps schema concepts to real-world entities.
    \item \textbf{Syntactic:} Evaluates naming conventions and string similarity.
    \item \textbf{Statistical:} Analyzes statistical distributions and inclusion relationships.
\end{itemize}

\subsection{Detailed Elaboration of Verifier}
% Algorithm~\ref{alg:conflict_resolution} outlines the detailed procedure of the Holistic Conflict Resolution strategy employed by the Verifier to enforce global structural consistency.
The Verifier enforces global consistency via the \textbf{Holistic Conflict Resolution} algorithm (see Algorithm~\ref{alg:conflict_resolution}), resolving multiple references and cyclic references to produce the final, structurally consistent FK set $\Phi$.

\begin{algorithm}[t]
\caption{Holistic Conflict Resolution}
\label{alg:conflict_resolution}
\begin{algorithmic}[1]
\REQUIRE Candidate FK Set $\Psi_{pos}$
\ENSURE Consistent FK Set $\Phi$

\STATE \textbf{Initialize} $\Phi \leftarrow \Psi_{pos}$

\STATE \textbf{// Stage 1: Resolve Multiple References}
\STATE \textit{// Ensure each source column references at most one target column}
\FOR{each source column $c_{src}$ referencing distinct target columns $\mathcal{C}_{targets}$ in $\Phi$}
    \IF{$|\mathcal{C}_{targets}| > 1$}
        \STATE \textit{// Select the single most plausible target column via reasoning}
        \STATE $c_{best} \leftarrow \text{Reasoning-Select}(c_{src}, \mathcal{C}_{targets})$
        \STATE Update $\Phi$ to retain only the reference $(c_{src}, c_{best})$ for $c_{src}$
    \ENDIF
\ENDFOR

\STATE \textbf{// Stage 2: Resolve Cyclic References}
\STATE \textit{// Iteratively eliminate cycles to ensure topological consistency}
\WHILE{Schema Topology Graph $G(\Phi)$ contains cycles}
    \STATE \textit{// Prioritize immediate violations for unambiguous resolution}
    \STATE $Cycle \leftarrow \text{FindShortestCycle}(G)$
    
    \STATE \textit{// Identify the least plausible link (edge) in the cycle}
    \STATE $e_{weak} \leftarrow \text{Reasoning-Weakest}(Cycle)$
    
    \STATE $\Phi \leftarrow \Phi \setminus \{e_{weak}\}$ 
    \COMMENT{Break the cycle}
\ENDWHILE

\RETURN $\Phi$
\end{algorithmic}
\end{algorithm}

\section{Dataset Details}
\label{dataset_details}
Table~\ref{tab:dataset_details} provides a detailed statistical overview of the datasets used in our experiments. These datasets span diverse domains, schema complexities, and scales, enabling a comprehensive evaluation of FK detection methods.

% \begin{table}[h]
% \centering
% \resizebox{\linewidth}{!}{
% \begin{tabular}{llllll}
% \hline\noalign{\smallskip}
% \textbf{Dataset} & \textbf{Type} & \textbf{Tables} & \textbf{FKs} & \textbf{Column Pairs} & \textbf{Domain} \\
% \noalign{\smallskip}\hline\noalign{\smallskip}
% TPC-H & Benchmark & 8 & 9 & 3,721 & Commerce \\
% Northwind & Synthetic & 13 & 11 & 7,744 & Retail \\
% TPC-E & Benchmark & 33 & 46 & 36,481 & Finance \\
% AdventureWorks & Synthetic & 71 & 91 & 236,196 & Manufacturing \\
% Musicbrainz & Real & 382 & 434 & 2,458,624 & Entertainment \\
% \noalign{\smallskip}\hline
% \end{tabular}
% }
% \caption{Detailed Statistics and Domains of Datasets.}
% \label{tab:dataset_details}
% \end{table}

\begin{table*}[t] % 改为 table* 以跨越双栏
\centering
\renewcommand{\arraystretch}{1.2} % 增加行高，让表格不那么拥挤
% 移除 resizebox，让字体保持文档默认大小，或者仅在必要时微调
\begin{tabular}{l l l l l l} 
\hline\noalign{\smallskip}
\textbf{Dataset} & \textbf{Type} & \textbf{Tables} & \textbf{FKs} & \textbf{Column Pairs} & \textbf{Domain} \\
\noalign{\smallskip}\hline\noalign{\smallskip}
TPC-H & Benchmark & 8 & 9 & 3,721 & Commerce \\
Northwind & Synthetic & 13 & 11 & 7,744 & Retail \\
TPC-E & Benchmark & 33 & 46 & 36,481 & Finance \\
AdventureWorks & Synthetic & 71 & 91 & 236,196 & Manufacturing \\
MusicBrainz & Real & 382 & 354 & 2,458,624 & Entertainment \\
\noalign{\smallskip}\hline
\end{tabular}
\caption{Detailed statistics and domains of datasets.}
\label{tab:dataset_details}
\end{table*}

% \begin{itemize}
%     \item \textbf{TPC-H}: A standard decision support benchmark widely used in the database community. It simulates a commercial environment with a relatively simple schema structure (8 tables) and sparse FK relationships, serving as a baseline for correctness verification.
%     \item \textbf{Northwind}: A classic synthetic dataset representing a small-scale retail business. With 13 tables, it provides a manageable yet representative schema for testing FK detection in typical inventory and sales scenarios.
%     \item \textbf{TPC-E}: An On-Line Transaction Processing (OLTP) benchmark that simulates a brokerage firm. It features a more complex schema (33 tables) compared to TPC-H, designed to represent modern financial transaction systems with richer integrity constraints. Notably, the schema is characterized by the frequent use of English abbreviations in its naming conventions.
%     \item \textbf{AdventureWorks}: A comprehensive synthetic database based on a fictional bicycle manufacturing company. It contains a large number of tables (71) and FKs, exhibiting complex schema patterns such as composite FKs, which pose significant challenges for detection algorithms.
%     \item \textbf{Musicbrainz}: A large-scale, real-world open music encyclopedia database. As the most complex dataset in our evaluation (382 tables and over 2.4 million column pairs), it tests the scalability and robustness of the methods in a noisy, real-world environment.
% \end{itemize}
\begin{itemize}
\item \textbf{TPC-H}: A standard decision-support benchmark widely adopted in the database community. It simulates a commercial environment with a relatively simple schema (8 tables) and sparse FK relationships, providing a baseline for correctness verification.
\item \textbf{Northwind}: A classic synthetic dataset representing a small retail business. With 13 tables, it offers a manageable yet representative schema for evaluating FK detection in typical inventory and sales scenarios.

\item \textbf{TPC-E}: An On-Line Transaction Processing (OLTP)~\citep{bernstein2009principles} benchmark modeling a brokerage firm. Its schema is more complex than TPC-H (33 tables) and includes richer integrity constraints. English abbreviations are frequently used in table and column names, introducing additional semantic challenges.

\item \textbf{AdventureWorks}: A synthetic database simulating a bicycle manufacturing company. Containing 71 tables and numerous FKs, it exhibits complex schema patterns, including composite FKs, which pose substantial challenges for FK detection algorithms.

\item \textbf{MusicBrainz}: A large-scale, real-world music encyclopedia database. With 382 tables and over 2.4 million column pairs, it is the most complex dataset in our evaluation, providing a testbed for both scalability and robustness in a noisy, heterogeneous environment.
\end{itemize}

\section{Reproducibility Statement}
\label{Reproducibility Statement}

\subsection{LLM-FK and LLM-based Baselines}
All experiments are conducted using \textbf{DeepSeek-R1} as the backbone LLM. To ensure deterministic results and reproducibility, we set the temperature parameter to $0$, while leaving all other hyperparameters at their default values.

For metadata extraction, we employ the \textbf{Metanome} profiling tool~\citep{papenbrock2015data} to identify structural constraints; specifically, we target \emph{MinUCCs} and utilize the \emph{Spider} algorithm~\citep{bauckmann2006efficiently} to detect single-column \emph{INDs}.

For LLM-based baselines:
\textbf{Chorus} is implemented using its official open-source code.
Detailed prompt designs for other LLM-based baselines, as well as for \textbf{LLM-FK}, are provided in Appendix~\ref{Prompts of LLM-FK and Case Study} and Appendix~\ref{Prompts of LLM-based Baselines}.

\subsection{Heuristic-based Baselines}
All heuristic-based baselines are faithfully re-implemented following the methodology described in their respective papers, with careful attention to hyperparameters and evaluation protocols:

\begin{itemize}
\item \textbf{MC-FK}: Since no default parameters are provided, we adopt the reported confidence scores for FK detection and empirically tune the decision threshold to maximize the F1-score on each dataset.
\item \textbf{Fast-FK}: Lacking explicit defaults, we assign reasonable weights to its constituent rules based on prior literature and preliminary tuning: column name similarity ($0.2$), column-table name similarity ($0.5$), cardinality ratio ($0.3$), and whether the referencing column is a unique key ($-0.5$).  

\item \textbf{HOPF}: Implemented strictly following the default configuration from the original work.  

\item \textbf{ML-FK} and \textbf{AutoSuggest}: We follow a leave-one-database-out training protocol proposed in the \textbf{ML-FK}. Specifically, for each target database, the labeled FK candidates from the remaining four datasets serve as training data. \textbf{LightGBM}~\citep{ke2017lightgbm} is employed as the classifier due to its efficiency and strong performance, with default hyperparameters used consistently across all experiments.  
\end{itemize}

\section{More Experiment Results and Analysis}
\label{sec:more_experiment}

\subsection{Generalizability to Different LLM Backbones}
\label{sec:generalizability_llms}

To demonstrate that LLM-FK's effectiveness does not rely on a particular LLM backbone, we evaluated our framework using DeepSeek-R1~\citep{guo2025deepseek}, Qwen3-Max~\citep{yang2025qwen3}, GPT-5-mini~\citep{singh2025openai}, and LLaMA3.3-70B-Instruct~\citep{grattafiori2024llama} without any model-specific fine-tuning. 

\paragraph{Generalizability of Candidate Pruning.} As shown in Table~\ref{tab:pruning_results}, the Profiler consistently achieves a substantial and reliable reduction in the candidate search space across all LLMs (e.g., from over 2.4 million to $\sim$2,355 pairs on MusicBrainz). Crucially, this reduction consistently preserves true positive matches, with only a single true FK being erroneously pruned across all trials—specifically by LLaMA3.3 on AdventureWorks. This highlights the high reliability and model-agnostic generalizability of our pruning methodology.

\paragraph{Generalizability of FK Detection.} Table~\ref{tab:final_f1_results} reports the F1-scores for FK detection. LLM-FK significantly outperforms the overall best-performing End-to-End baselines specifically on complex schemas such as TPC-E, AdventureWorks, and MusicBrainz, demonstrating the generalizability of our multi-agent collaborative approach across different underlying LLMs. Among the evaluated models, LLaMA 3.3 exhibits relatively lower performance, likely attributable to differences in inherent reasoning capabilities for complex logical constraints. Nevertheless, these results still consistently surpass the vast majority of traditional heuristic-based methods (see Table~\ref{tab:overall_performance}) and End-to-End baselines, further supporting the effectiveness of our LLM-FK. Finally, the marginally lower F1-scores observed on smaller datasets like TPC-H are an artifact of the constrained schema size; in practice, only a few FKs was misclassified, which disproportionately skews the metric rather than indicating a fundamental limitation.

\begin{table*}[htbp]
    \centering
    \small
    \begin{tabular}{l r c c c c c c c c}
        \toprule
        \textbf{Dataset} & \textbf{Original Pairs} & \multicolumn{2}{c}{\textbf{DeepSeek-R1}} & \multicolumn{2}{c}{\textbf{Qwen3-Max}} & \multicolumn{2}{c}{\textbf{LLaMA3.3}} & \multicolumn{2}{c}{\textbf{GPT-5-mini}} \\
        \cmidrule(lr){3-4} \cmidrule(lr){5-6} \cmidrule(lr){7-8} \cmidrule(lr){9-10}
        & & \textbf{Cand.} & \textbf{Err.} & \textbf{Cand.} & \textbf{Err.} & \textbf{Cand.} & \textbf{Err.} & \textbf{Cand.} & \textbf{Err.} \\
        \midrule
        TPC-H          & 3,721     & 49    & 0 & 49    & 0 & 47    & 0 & 49    & 0 \\
        Northwind      & 7,744     & 60    & 0 & 60    & 0 & 60    & 0 & 60    & 0 \\
        TPC-E          & 36,481    & 194   & 0 & 194   & 0 & 195   & 0 & 192   & 0 \\
        AdventureWorks & 236,196   & 2,034 & 0 & 2,057 & 0 & 1,987 & 1 & 2,024 & 0 \\
        MusicBrainz    & 2,458,624 & 2,306 & 0 & 2,355 & 0 & 2,355 & 0 & 2,355 & 0 \\
        \bottomrule
    \end{tabular}
    \caption{The Profiler candidate pruning results across different LLMs. Each entry shows the number of candidates (Cand.) and the number of erroneous prunings (Err.), where erroneous pruning indicates FKs that are incorrectly removed during the pruning process.}
    \label{tab:pruning_results}
\end{table*}

\begin{table*}[htbp]
    \centering
    \small
    \begin{tabular}{l l c c c c c}
        \toprule
        \textbf{LLM} & \textbf{Method} & \textbf{TPC-H} & \textbf{Northwind} & \textbf{TPC-E} & \textbf{AdventureWorks} & \textbf{MusicBrainz} \\
        \midrule
        \multirow{2}{*}{DeepSeek-R1} & End-to-End & 0.90 & 1.00 & 0.69 & 0.73 & 0.75 \\
                                     & LLM-FK (Ours) & \textbf{1.00} & \textbf{1.00} & \textbf{0.93} & \textbf{0.94} & \textbf{0.95} \\
        \midrule
        \multirow{2}{*}{Qwen3-Max}   & End-to-End & \textbf{0.90} & 0.92 & 0.58 & 0.52 & 0.75 \\
                                     & LLM-FK (Ours) & 0.78 & \textbf{1.00} & \textbf{0.92} & \textbf{0.92} & \textbf{0.96} \\
        \midrule
        \multirow{2}{*}{LLaMA3.3}    & End-to-End & \textbf{0.80} & 0.95 & 0.61 & 0.66 & 0.74 \\
                                     & LLM-FK (Ours) & 0.70 & \textbf{0.96} & \textbf{0.85} & \textbf{0.79} & \textbf{0.86} \\
        \midrule
        \multirow{2}{*}{GPT-5-mini}  & End-to-End & \textbf{0.90} & 0.95 & 0.67 & 0.79 & 0.80 \\
                                     & LLM-FK (Ours) & 0.78 & \textbf{1.00} & \textbf{0.93} & \textbf{0.95} & \textbf{0.93} \\
        \bottomrule
    \end{tabular}
    \caption{F1-score comparison for FK detection across different LLMs. Best results per LLM are bolded.}
    \label{tab:final_f1_results}
\end{table*}

\subsection{Computational Cost Analysis}
\label{sec:computational_cost}

Following the preliminary search space analysis in Section \ref{Analysis of Candidate Pruning Efficiency}, we detail the runtime overhead and computational expenditures associated with LLM calls. We conduct a comprehensive analysis of API calls, token usage, and monetary costs using Qwen3-Max as the backend LLM. The pricing is based on Qwen3-Max, with a rate of 2.5 CNY for inputs up to 32K tokens and 10 CNY for outputs up to 32K tokens. 

\paragraph{Overall Comparison.} To evaluate the LLM invocation frequency and associated expenses, we compare our method (LLM-FK) against an End-to-End baseline. This baseline identifies FKs between a pair of tables in a single pass and serves as the most effective and efficient baseline in our study. The overall comparison is summarized in Table~\ref{tab:overall_cost}. For smaller schemas such as TPC-H, LLM-FK exhibits a slightly lower F1-score (0.78 compared to 0.90) and incurs marginally higher costs. However, this performance gap corresponds to merely a few misclassified FKs, which is magnified by the small scale of the dataset. For larger datasets, including TPC-E and AdventureWorks, LLM-FK requires higher API costs and token consumption. Nevertheless, this increased expenditure represents a favorable trade-off, as it yields substantial improvements in accuracy and better performance in terms of API calls. Most notably, when applied to massive and complex schemas like MusicBrainz, LLM-FK not only achieves the highest accuracy (0.96) but also drastically reduces monetary costs, demonstrating its superior scalability and efficiency in large-scale environments.

% \paragraph{Internal Breakdown.} To further elucidate the sources of the computational overhead, we provide a detailed breakdown across the internal agents within our multi-agent framework: Profiler, Interpreter, Refiner, and Verifier. As detailed in Table~\ref{tab:agent_breakdown}, the analysis clearly indicates that the primary time and cost expenditures are concentrated within the Refiner module.

\paragraph{Internal Breakdown.} To further elucidate the sources of the computational overhead, we provide a detailed breakdown across the internal agents within our multi-agent framework: Profiler, Interpreter, Refiner, and Verifier. As detailed in Table~\ref{tab:agent_breakdown}, the analysis clearly indicates that the primary time and cost expenditures are concentrated within the Refiner module. Conversely, the Verifier incurs minimal overhead because its input is drastically pruned by the Refiner, such as reducing 2,355 candidates down to merely 92 conflicts in the \texttt{MusicBrainz} dataset. Furthermore, these remaining conflicts naturally form tightly clustered, localized semantic subgraphs; notably, all cyclic references encountered across our experiments are strictly length-2 (see Appendix~\ref{sec:example_holistic} for a typical example). By iteratively prioritizing the resolution of these shortest cycles, the Verifier processes highly isolated, immediate neighborhood conflicts rather than complex, cascading chains, thereby requiring significantly fewer computational overhead to globally resolve.

% Additionally, the execution times reported in this study reflect sequential LLM queries. In practical deployments, the framework's latency can be substantially reduced by processing queries concurrently via asynchronous API calls.

\subsection{Robustness in the Absence of UCC Metadata}
\label{sec:robustness_no_ucc}

To evaluate the pruning effectiveness of our method under empty-table conditions (where UCCs are unavailable), we conducted an experiment to verify its robustness. As summarized in Table \ref{tab:empty_table_robustness}, the results demonstrate consistently excellent pruning performance across all five benchmark datasets. The volume of retained candidates remains highly comparable to that of the original data-rich environment, and only a single true FK (in TPC-E) was erroneously pruned. This high consistency is achieved because, as illustrated in Table \ref{tab:uk_selection_rules}, our method leverages LLMs to comprehensively reason across multiple dimensions—such as column ordinality, naming conventions, and semantic alignment. This multi-dimensional reasoning allows our framework to infer plausible referenced unique keys directly from the schema, effectively compensating for the lack of UCCs and ensuring robustness even in completely empty-table environments. Furthermore, it is worth noting that in practical database operations, completely empty tables are often temporary or staging tables that rarely require complex structural analysis.

\begin{table}[htbp]
    \centering
    \footnotesize % 使用比正文小一号的字体，保证单栏能放下且清晰
    \setlength{\tabcolsep}{4pt} % 缩小列间距 (默认通常是6pt)，非常关键的省空间技巧
    \begin{tabular}{l r c c c c}
        \toprule
        \textbf{Dataset} & \textbf{Pairs} & \multicolumn{2}{c}{\textbf{Original}} & \multicolumn{2}{c}{\textbf{Empty}} \\
        \cmidrule(lr){3-4} \cmidrule(lr){5-6}
        & & \textbf{Cand.} & \textbf{Err.} & \textbf{Cand.} & \textbf{Err.} \\
        \midrule
        TPC-H          & 3,721     & 49    & 0 & 49    & 0 \\
        Northwind      & 7,744     & 60    & 0 & 69    & 0 \\
        TPC-E          & 36,481    & 194   & 0 & 156   & 1 \\
        AdventureWorks & 236,196   & 2,034 & 0 & 2,127 & 0 \\
        MusicBrainz    & 2,458,624 & 2,306 & 0 & 2,138 & 0 \\
        \bottomrule
    \end{tabular}
    \caption{Pruning results in original versus empty tables. Cand. denotes retained candidates; Err. denotes mistakenly pruned true FKs.}
    \label{tab:empty_table_robustness}
\end{table}

\subsection{Example Analysis of Unique-Key-Driven Schema Decomposition}
\label{sec:example_analysis_decomposition}

\paragraph{Column-Level Processing vs. Table-Level Processing.}
Prior methods input entire table pairs to LLMs, requesting all FKs in a single pass. This often leads to attention dispersion and hallucinations, particularly when multiple FKs exist between the same tables.

For example, in TPC-E, the table \texttt{\detokenize{holding_history}} references \texttt{\detokenize{trade}} through two distinct FKs: \texttt{\detokenize{hh_h_t_id}} $\to$ \texttt{\detokenize{t_id}} and \texttt{\detokenize{hh_t_id}} $\to$ \texttt{\detokenize{t_id}}. In our experiments, table-level methods typically detect only one FK. In contrast, even without Self-Augmented Domain Knowledge Injection or Multi-Perspective CoT Reasoning, LLM-FK successfully identifies both FKs, attributed to our column-pair decomposition. By decomposing schemas into atomic units, we reduce task complexity invariant to schema size, allowing focused reasoning for each candidate pair and maintaining tractability in large schemas.

\paragraph{Unique Key-Based vs. Primary Key-Based Pruning.}
Our unique key-based pruning is more robust than relying solely on declared PKs. In complex schemas, PKs often form supersets of MinUCCs, introducing unnecessary redundancy. For instance, in \texttt{\detokenize{production_productproductphoto}} (AdventureWorks), the PK is composite (\texttt{\detokenize{ProductID}} + \texttt{\detokenize{ProductPhotoID}}), whereas \texttt{\detokenize{ProductID}} alone constitutes a MinUCC and serves as the target for all referencing FKs. 
% All FKs reference \texttt{\detokenize{ProductID}}. 
If pruning relied on the overly composite PK, irrelevant candidate pairs involving \texttt{\detokenize{ProductPhotoID}} would persist, increasing computational cost. In contrast, unique key-based pruning efficiently narrows the search space while preserving FK integrity.

\subsection{Example Analysis of Self-Augmented Domain Knowledge Injection}
\label{sec:example_domain_knowledge}
We analyze the \texttt{\detokenize{holding_summary.hs_s_symb}} and \texttt{\detokenize{daily_market.dm_s_symb}} candidate column pair from TPC-E. Both columns share the suffix \texttt{\detokenize{s_symb}}    (Security Symbol) but serve distinct roles: \texttt{\detokenize{daily_market}} tracks daily stock data, whereas \texttt{\detokenize{holding_summary}} aggregates customer holdings.

Without domain knowledge, the LLM fails to decipher the semantics of these abbreviations, resulting in an erroneous reliance on syntactic similarity and mistakenly predicting an FK. With our Interpreter, a global domain context identifies the database as a “Financial Trading System” and distinguishes “Market Data” from “Account Holdings.” This knowledge guides the Refiner to reject the candidate, illustrating that self-augmented domain knowledge resolves semantic ambiguities effectively.

\subsection{Example Analysis of Multi-Perspective CoT Reasoning}
\label{sec:example_multi_perspective_cot}
Beyond ensuring the robustness of LLM-FK in scenarios with missing data or ambiguous semantics, the Multi-Perspective CoT Reasoning leverages sufficient information embedded in structural representations and performs deep semantic reasoning to address ambiguous or inconsistent signals. For instance,  the FK \texttt{series\_annotation.annotation} $\to$ \texttt{annotation.id} in MusicBrainz involves columns with numerical values in different formats (scientific vs. integer), where the former is formalized as scientific notation due to the large magnitude of values in the leading rows, while the latter remains in integer format. Without multi-perspective reasoning, the LLM misinterprets scientific notation as string values, rejecting the FK due to mismatches in both data type and value range within the leading rows. LLM-FK overcomes this by:
\begin{itemize}
\item Injecting statistical metadata (data types, value ranges) to provide sufficient information, preventing misinterpretation.

% \item Cross-validating syntactic, semantic, and data perspectives to ensure robust inference despite format discrepancies.
\item Cross-validating syntactic, semantic and statistical perspectives to ensure robust inference transcending superficial judgments based solely on format discrepancies or value range mismatches.
\end{itemize}

% \subsection{Example Analysis of Holistic Conflict Resolution}
% \label{sec:example_holistic}
% In MusicBrainz, \texttt{l\_artist\_recording.entity0} is erroneously linked to both \texttt{artist\_meta.id} and \texttt{artist.id}, forming a cyclic reference. Without Holistic Conflict Resolution, these conflicts remain unresolved.

% By constructing a global schema topology, LLM-FK identifies \texttt{artist} as the core entity and \texttt{artist\_meta} as auxiliary metadata. Hierarchical reasoning eliminates redundant references and ensures global structural and semantic consistency, outperforming local confidence-based disambiguation.

\subsection{Example Analysis of Holistic Conflict Resolution}
\label{sec:example_holistic}

In the MusicBrainz, complex conflicts arise involving \texttt{l\_artist\_recording}, \texttt{artist}, and \texttt{artist\_meta} within the outputs of the Refiner. 
Specifically, \texttt{l\_artist\_recording.entity0} is ambiguously identified as an FK referencing both \texttt{artist.id} and \texttt{artist\_meta.id}.
Exacerbating this issue, a direct cyclic reference exists between \texttt{artist} and \texttt{artist\_meta} (i.e., they reference each other), which obscures the true hierarchical relationship. 
Such structural conflicts are highly typical rather than random occurrences. Crucially, the Refiner’s semantic-based reasoning constrains these highly correlated candidates into tightly clustered, semantically related subgraphs.
% Rather than occurring randomly, this structural conflict forms a semantically related subgraph, as the Refiner's semantic-based reasoning typically constrains these highly correlated entities together.

By constructing the global schema topology and collectively analyzing the candidates within the conflicting subgraphs, our Holistic Conflict Resolution Strategy identifies that \texttt{artist} serves as the semantic core entity, whereas \texttt{artist\_meta} acts as auxiliary metadata. Consequently, we resolve the conflicts by retaining the canonical references to \texttt{artist.id} and eliminating the spurious candidates referencing \texttt{artist\_meta.id}. 
This decision improves upon local validation methods that rely on assigning confidence scores to column pairs. While such scores are often inaccurate due to isolated reasoning, our construction of a global schema topology facilitates a direct semantic comparison to capture hierarchical relationships. This holistic reasoning process eliminates conflicting references and ensures that the final alignment is both semantically and structurally consistent at the global schema level.

\begin{table*}
\centering
\resizebox{\textwidth}{!}{ 
\begin{tabular}{llccccccc}
\toprule
\multirow{2}{*}{\textbf{Dataset}} & \multirow{2}{*}{\textbf{Method}} & \multirow{2}{*}{\textbf{F1-score}} & \multirow{2}{*}{\textbf{API Calls}} & \multicolumn{2}{c}{\textbf{Tokens}} & \multicolumn{3}{c}{\textbf{Cost (¥)}} \\
\cmidrule(lr){5-6} \cmidrule(lr){7-9}
& & & & \textbf{In} & \textbf{Out} & \textbf{In} & \textbf{Out} & \textbf{Total} \\
\midrule
\multirow{2}{*}{TPC-H} 
& End-to-End & \textbf{0.90} & \textbf{36} & \textbf{0.0351} & \textbf{0.0006} & \textbf{0.088} & \textbf{0.006} & \textbf{0.094} \\
& LLM-FK & 0.78 & 60 & 0.1221 & 0.0011 & 0.305 & 0.011 & 0.316 \\
\midrule
\multirow{2}{*}{Northwind} 
& End-to-End & 0.92 & \textbf{66} & \textbf{0.0617} & \textbf{0.0010} & \textbf{0.154} & \textbf{0.010} & \textbf{0.164} \\
& LLM-FK & \textbf{1.00} & 71 & 0.1312 & 0.0012 & 0.328 & 0.012 & 0.339 \\
\midrule
\multirow{2}{*}{TPC-E} 
& End-to-End & 0.58 & 528 & \textbf{0.4496} & 0.0066 & \textbf{1.124} & 0.066 & \textbf{1.190} \\
& LLM-FK & \textbf{0.92} & \textbf{245} & 0.5342 & \textbf{0.0031} & 1.335 & \textbf{0.032} & 1.367 \\
\midrule
\multirow{2}{*}{AdventureWorks} 
& End-to-End & 0.52 & 2485 & \textbf{2.6313} & 0.0346 & \textbf{6.578} & 0.346 & \textbf{6.924} \\
& LLM-FK & \textbf{0.92} & \textbf{2142} & 5.5458 & \textbf{0.0224} & 13.864 & \textbf{0.223} & 14.087 \\
\midrule
\multirow{2}{*}{MusicBrainz} 
& End-to-End & 0.75 & 28203 & 25.9330 & 0.3653 & 64.832 & 3.653 & 68.485 \\
& LLM-FK & \textbf{0.96} & \textbf{2588} & \textbf{6.3064} & \textbf{0.0264} & \textbf{15.766} & \textbf{0.264} & \textbf{16.029} \\
\bottomrule
\end{tabular}
}
\caption{Overall computational cost comparison between the End-to-End baseline and LLM-FK across different datasets.}
\label{tab:overall_cost}
\end{table*}

\begin{table*}[htbp]
\centering
\begin{tabular}{llcccccc}
\toprule
\multirow{2}{*}{\textbf{Dataset}} & \multirow{2}{*}{\textbf{Agent}} & \multirow{2}{*}{\textbf{API Calls}} & \multicolumn{2}{c}{\textbf{Tokens}} & \multicolumn{3}{c}{\textbf{Cost (¥)}} \\
\cmidrule(lr){4-5} \cmidrule(lr){6-8}
& & & \textbf{In} & \textbf{Out} & \textbf{In} & \textbf{Out} & \textbf{Total} \\
\midrule
\multirow{4}{*}{TPC-H} 
& Profiler & 8 & 0.0149 & 0.0001 & 0.037 & 0.001 & 0.038 \\
& Interpreter & 1 & 0.0003 & 0.0005 & 0.001 & 0.005 & 0.006 \\
& Refiner & \textbf{49} & 0.1004 & 0.0005 & 0.251 & 0.005 & \textbf{0.256} \\
& Verifier & 2 & 0.0064 & 0.0000 & 0.016 & 0.000 & 0.017 \\
\midrule
\multirow{4}{*}{Northwind} 
& Profiler & 9 & 0.0067 & 0.0001 & 0.017 & 0.001 & 0.018 \\
& Interpreter & 1 & 0.0004 & 0.0005 & 0.001 & 0.005 & 0.005 \\
& Refiner & \textbf{60} & 0.1211 & 0.0006 & 0.303 & 0.006 & \textbf{0.309} \\
& Verifier & 1 & 0.0030 & 0.0000 & 0.007 & 0.000 & 0.008 \\
\midrule
\multirow{4}{*}{TPC-E} 
& Profiler & 20 & 0.0140 & 0.0002 & 0.035 & 0.002 & 0.037 \\
& Interpreter & 1 & 0.0004 & 0.0006 & 0.001 & 0.006 & 0.007 \\
& Refiner & \textbf{194} & 0.4165 & 0.0019 & 1.041 & 0.019 & \textbf{1.061} \\
& Verifier & 30 & 0.1032 & 0.0004 & 0.258 & 0.004 & 0.262 \\
\midrule
\multirow{4}{*}{AdventureWorks} 
& Profiler & 60 & 0.0433 & 0.0006 & 0.108 & 0.006 & 0.114 \\
& Interpreter & 1 & 0.0007 & 0.0009 & 0.002 & 0.009 & 0.011 \\
& Refiner & \textbf{2057} & 5.3956 & 0.0205 & 13.489 & 0.205 & \textbf{13.694} \\
& Verifier & 24 & 0.1061 & 0.0004 & 0.265 & 0.004 & 0.269 \\
\midrule
\multirow{4}{*}{MusicBrainz} 
& Profiler & 140 & 0.0967 & 0.0014 & 0.242 & 0.014 & 0.256 \\
& Interpreter & 1 & 0.0014 & 0.0011 & 0.004 & 0.011 & 0.015 \\
& Refiner & \textbf{2355} & 5.8799 & 0.0228 & 14.700 & 0.228 & \textbf{14.927} \\
& Verifier & 92 & 0.3284 & 0.0011 & 0.821 & 0.011 & 0.832 \\
\bottomrule
\end{tabular}
\caption{Detailed breakdown of computational overhead across internal agents in the LLM-FK framework.}
\label{tab:agent_breakdown}
\end{table*}

\begin{algorithm*}
\caption{The Overall Pipeline of LLM-FK}
\label{alg:llm_fk_pipeline}
\begin{algorithmic}[1]
\REQUIRE Database $\mathcal{D}$, Instruction Set $\mathcal{I} = \{\mathcal{I}_{\mathcal{A}_{p}}, \mathcal{I}_{\mathcal{A}_{i}}, \mathcal{I}_{\mathcal{A}_{r}}, \mathcal{I}_{\mathcal{A}_{v}}\}$
\ENSURE The set of FKs $\Phi$

\STATE \textbf{// Stage 1: Profiler ($\mathcal{A}_{p}$) - Profiling \& Unique-Key-Driven Schema Decomposition}
\STATE $\mathcal{J}_{raw} \leftarrow \text{ExtractINDs}(\mathcal{D})$ \COMMENT{Profiler invokes tool to extract raw INDs}
\STATE $\mathcal{U} \leftarrow \text{ExtractMinUCCs}(\mathcal{D})$ \COMMENT{Profiler invokes tool to extract MinUCCs}
\STATE $\mathcal{J} \leftarrow \text{FilterByType}(\mathcal{J}_{raw})$ \COMMENT{Profiler invokes tool to filter INDs by data types}
% \STATE $\mathcal{J} \leftarrow \text{FilterByType}(\mathcal{J}_{raw})$ \COMMENT{Pre-filtering based on data types}
\STATE $\mathcal{K} \leftarrow \emptyset$ 
\COMMENT{Initialize the set of selected unique keys}
\STATE $\mathcal{T}_{ref} \leftarrow \text{GetReferencedTables}(\mathcal{J})$
\FOR{each table $T \in \mathcal{T}_{ref}$}
    \STATE $k \leftarrow \pi(T, \mathcal{U}_T, \mathcal{I}_{\mathcal{A}_{p}})$ 
    \COMMENT{Profiler selects the most plausible referenced unique key}
    \STATE $\mathcal{K} \leftarrow \mathcal{K} \cup \{k\}$
\ENDFOR
\STATE $\Psi_{cand} \leftarrow \{(c_f, c_p) \in \mathcal{J} \mid c_p \in \mathcal{K}\}$ 
\COMMENT{Profiler prunes $\mathcal{J}$ using selected unique keys}

\STATE \textbf{// Stage 2: Interpreter ($\mathcal{A}_{i}$) - Self-Augmented Domain Knowledge Injection}
\STATE $\mathcal{N}_{\Psi_{cand}} \leftarrow \text{ExtractTableNames}(\Psi_{cand})$ 
\COMMENT{Interpreter invokes tool to extract table names}
\STATE $\mathcal{K}_{dom} \leftarrow \pi(\mathcal{N}_{\Psi_{cand}}, \mathcal{I}_{\mathcal{A}_{i}})$ 
\COMMENT{Interpreter constructs self-augmented domain knowledge}

\STATE \textbf{// Stage 3: Refiner ($\mathcal{A}_{r}$) - Multi-Perspective CoT Reasoning}
\STATE $\Psi_{pos} \leftarrow \emptyset$ 
\COMMENT{Initialize local FK result set}
\FOR{each candidate $\psi \in \Psi_{cand}$ \textbf{in parallel}}
    \STATE $S_\psi \leftarrow \text{GetStructuralRepresentative}(\psi, \mathcal{D})$ 
    \COMMENT{Refiner formulates SQL queries guided by Multi-Perspective CoT to derive structural representations}
    \STATE $p_{\psi} \leftarrow \pi(\psi, S_\psi, \mathcal{K}_{dom}, \mathcal{I}_{\mathcal{A}_{r}})$ 
    \COMMENT{Refiner executes binary validation ($p_{\psi} \in \{0, 1\}$)}
    \IF{$p_{\psi} = 1$}
        \STATE $\Psi_{pos} \leftarrow \Psi_{pos} \cup \{\psi\}$
    \ENDIF
\ENDFOR

\STATE \textbf{// Stage 4: Verifier ($\mathcal{A}_{v}$) - Holistic Conflict Resolution}
\STATE $\mathcal{G} \leftarrow \text{ConstructGraph}(\Psi_{pos})$ 
\COMMENT{Verifier constructs schema topology graph}

\STATE \textit{// Phase 4.1: Resolve Multiple References First}
\STATE $\mathbf{C}_{multi} \leftarrow \text{DetectMultiRefConflicts}(\mathcal{G})$ 
\COMMENT{Verifier invokes tool to detect multiple references}
\FOR{each conflict set $\Psi_{conflict} \in \mathbf{C}_{multi}$}
    \STATE $S_{\Psi_{conflict}} \leftarrow \text{GetStructuralRepresentative}(\Psi_{conflict}, \mathcal{D})$ 
    \COMMENT{Verifier formulates SQL queries guided by Multi-Perspective CoT to derive structural representations}
    \STATE $\Psi_{resolved} \leftarrow \pi(\Psi_{conflict}, S_{\Psi_{conflict}}, \mathcal{K}_{dom}, \mathcal{I}_{\mathcal{A}_{v}})$ 
    \COMMENT{Verifier resolves conflicts by retaining the most plausible candidate FK}
    \STATE $\Psi_{pos} \leftarrow (\Psi_{pos} \setminus \Psi_{conflict}) \cup \Psi_{resolved}$
\ENDFOR

\STATE \textit{// Phase 4.2: Iteratively Resolve Cyclic References}
\WHILE{TRUE}
    \STATE $\mathcal{G} \leftarrow \text{ConstructGraph}(\Psi_{pos})$ \COMMENT{Verifier constructs the graph with updated edges}
    \STATE $\mathbf{C}_{cycle} \leftarrow \text{DetectCyclicConflicts}(\mathcal{G})$ 
    \COMMENT{Verifier invokes tool to detect cyclic references}
    \IF{$\mathbf{C}_{cycle} = \emptyset$}
        \STATE \textbf{break} 
        \COMMENT{Terminate if no cycles remain}
    \ENDIF
    \STATE Select the shortest conflict set $\Psi_{conflict} \in \mathbf{C}_{cycle}$ 
    \COMMENT{Verifier prioritizes the shortest cycle for resolution}
    \STATE $S_{\Psi_{conflict}} \leftarrow \text{GetStructuralRepresentative}(\Psi_{conflict}, \mathcal{D})$
    \STATE $\Psi_{resolved} \leftarrow \pi(\Psi_{conflict}, S_{\Psi_{conflict}}, \mathcal{K}_{dom}, \mathcal{I}_{\mathcal{A}_{v}})$ 
    \COMMENT{Verifier removes the least plausible candidate FK to break the cycle}
    \STATE $\Psi_{pos} \leftarrow (\Psi_{pos} \setminus \Psi_{conflict}) \cup \Psi_{resolved}$
\ENDWHILE

\STATE $\Phi \leftarrow \Psi_{pos}$
\RETURN $\Phi$
\end{algorithmic}
\end{algorithm*}

\section{Prompts of LLM-FK and Case Study}
\label{Prompts of LLM-FK and Case Study}
% In this appendix, we present detailed prompt examples for the core strategies employed by the Profiler, Interpreter, Refiner, and Verifier. We utilize the sample database introduced in Figure \ref{fig:2} as our primary case study. Figure \ref{fig:1} illustrates the advantages of LLM-FK over heuristic-based methods, Figure \ref{fig:2} depicts the overall multi-agent collaboration workflow for FK detection, and Figure \ref{fig:interpreter_workflow} details the Self-Augmented Domain Knowledge Injection. Correspondingly, Figures \ref{fig:6} through \ref{fig:10} display the specific prompts and their corresponding outputs across different stages of the pipeline.
In this appendix, we provide detailed examples of the prompts used for each core agent in LLM-FK: Profiler, Interpreter, Refiner, and Verifier. These examples illustrate how the multi-agent framework operationalizes FK detection.

We use the sample database introduced in Figure~\ref{fig:2} as the running case study, which also illustrates the overall multi-agent workflow.
Figure~\ref{fig:1} illustrates the limitations of heuristic-based methods and the effectiveness of LLMs in semantic reasoning for FK detection, while also highlighting the accompanying challenges.
Figure~\ref{fig:interpreter_workflow} specifically details the Self-Augmented Domain Knowledge Injection performed by the Interpreter.

\begin{figure*} 
  \centering
  \includegraphics[width=\linewidth]{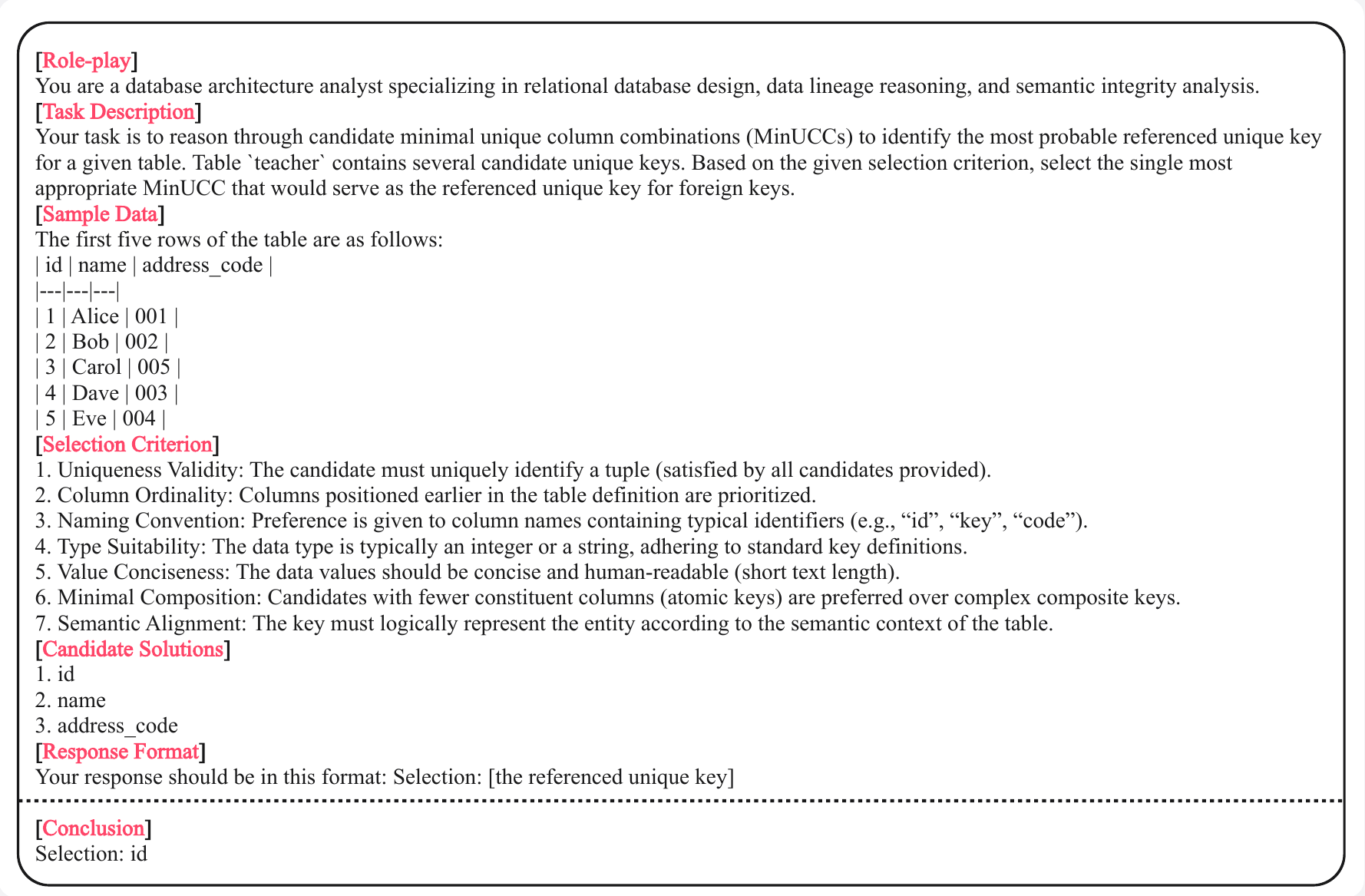}
  \caption{Prompt of the Profile for unique key selection.}
  \label{fig:6}
\end{figure*}

\begin{figure*} 
  \centering
  \includegraphics[width=\linewidth]{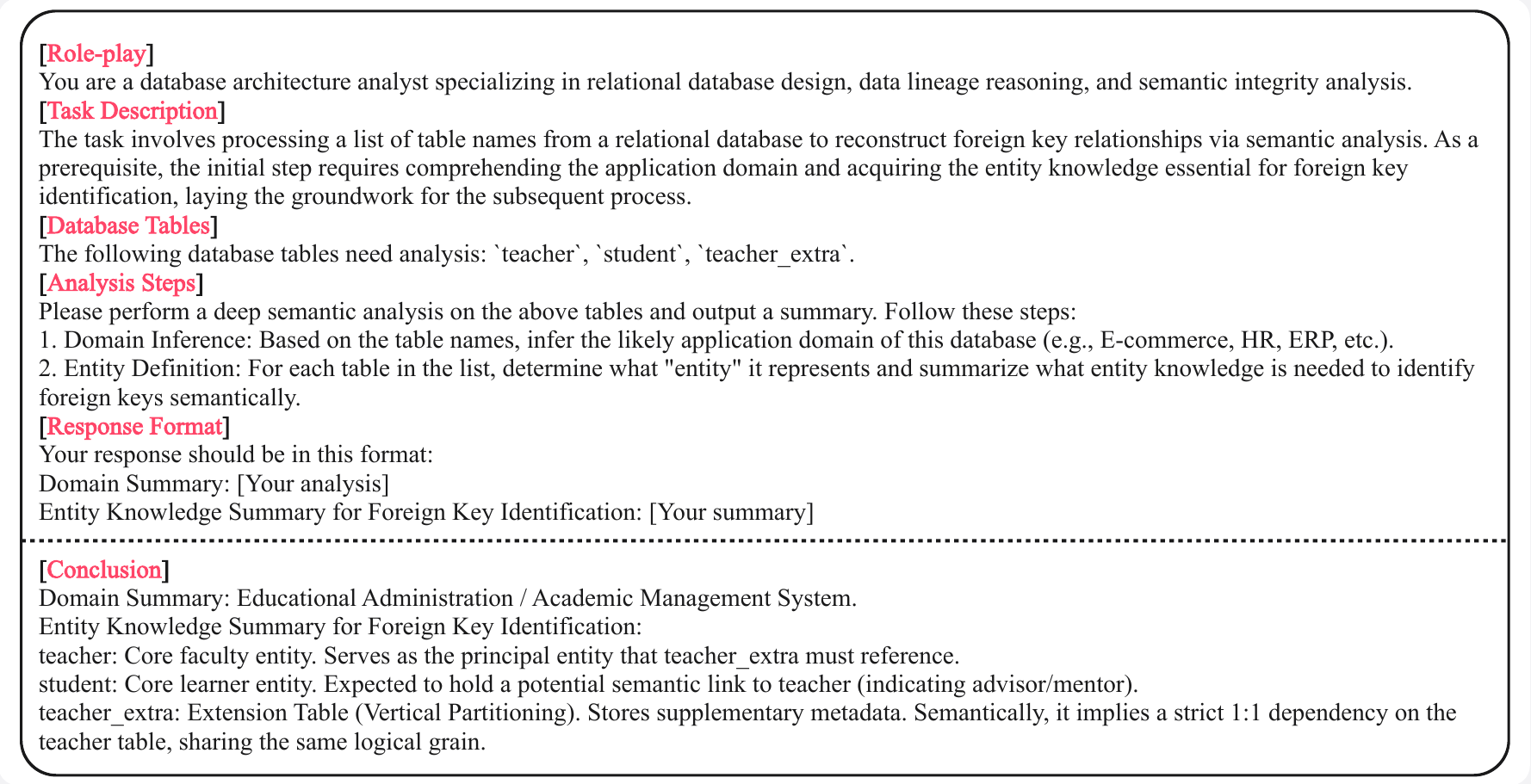}
  \caption{Prompt of the Interpreter for domain knowledge generation.}
  \label{fig:7}
\end{figure*}

\begin{figure*} 
  \centering
  \includegraphics[width=\linewidth]{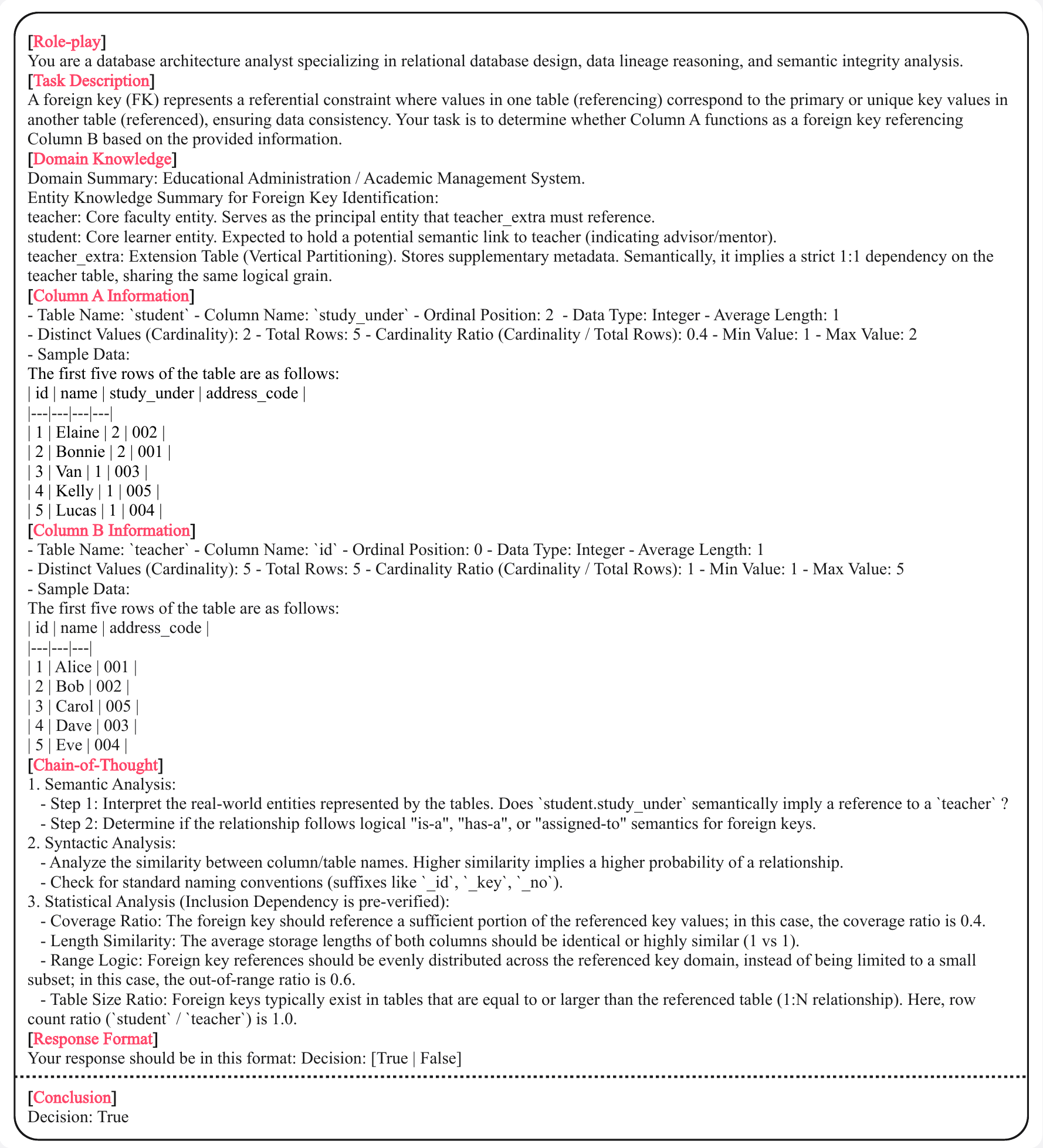}
  \caption{Prompt of the Refiner for each FK candidate pair validation.}
  \label{fig:8}
\end{figure*}

\begin{figure*} 
  \centering
  \includegraphics[width=\linewidth]{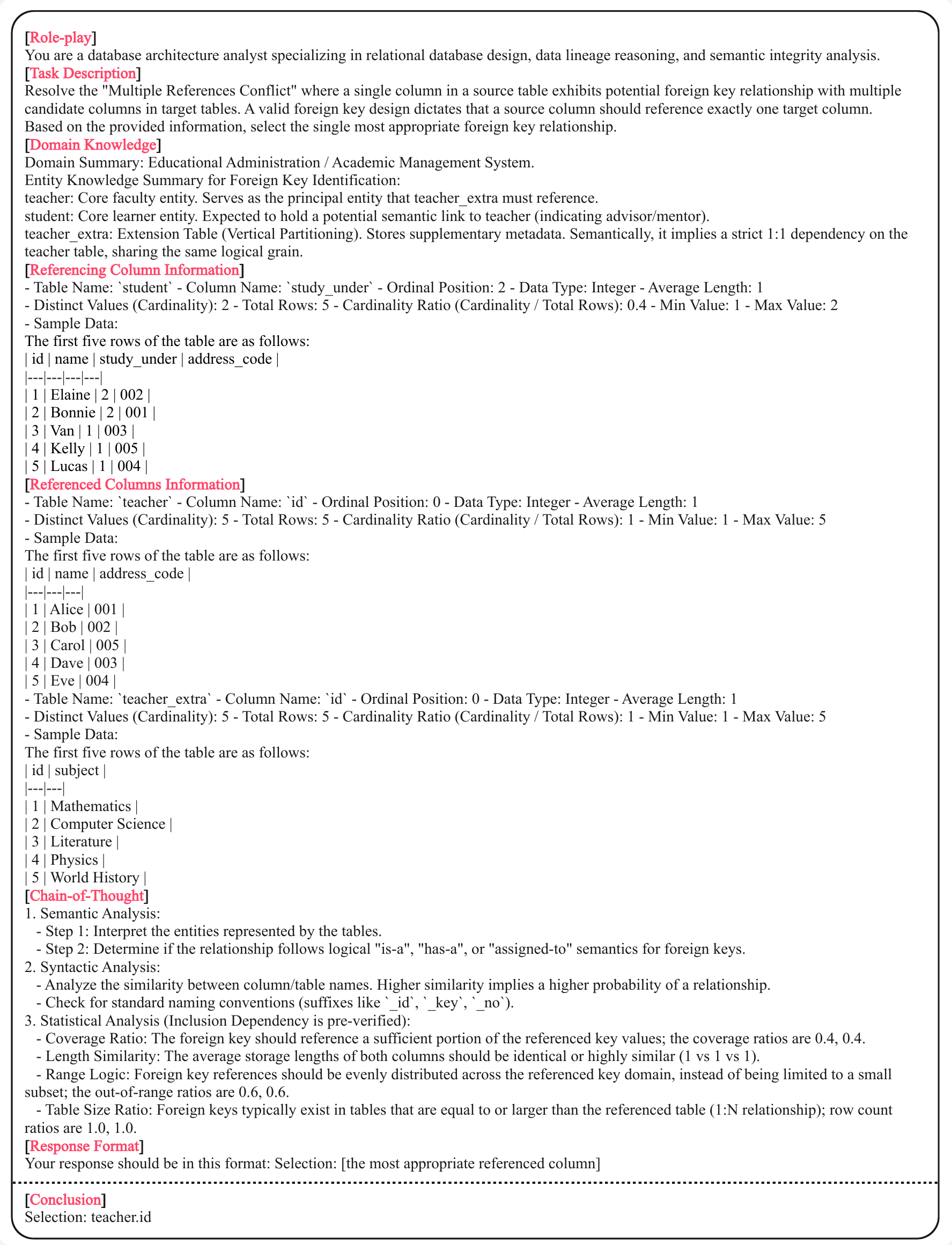}
  \caption{Prompt of the Verifier for multiple references resolution.}
  \label{fig:9}
\end{figure*}

\begin{figure*} 
  \centering
  \includegraphics[width=\linewidth]{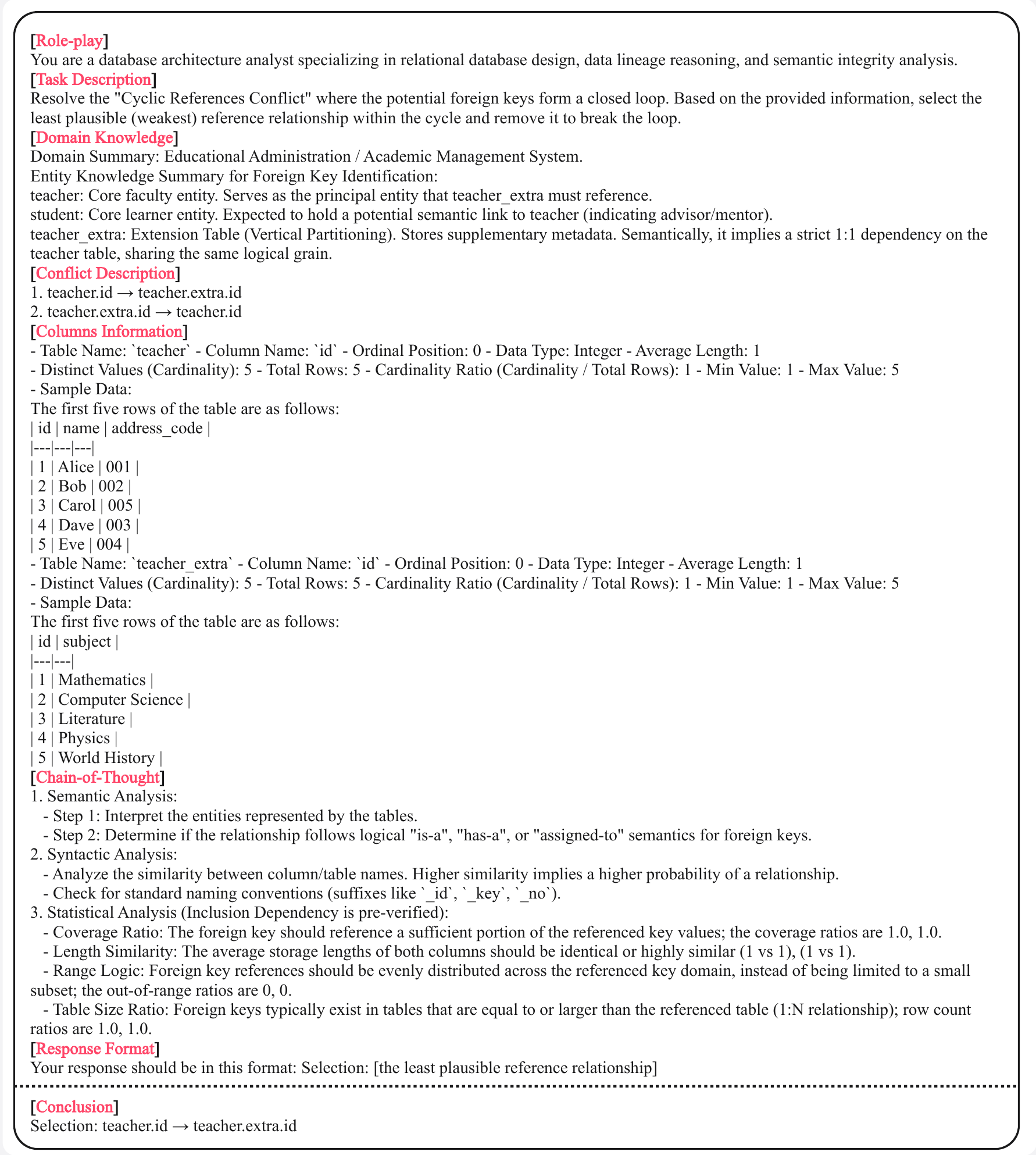}
  \caption{Prompt of the Verifier for cyclic references resolution.}
  \label{fig:10}
\end{figure*}

Figures~\ref{fig:6}--\ref{fig:10} show the specific prompts and their corresponding outputs across different stages of the pipeline. Specifically, these example prompts are directly populated with the structural representations, representing the execution outcomes of the SQL queries.

\section{Prompts of LLM-based Baselines}
\label{Prompts of LLM-based Baselines}
\begin{figure*} 
  \centering
  \includegraphics[width=\linewidth]{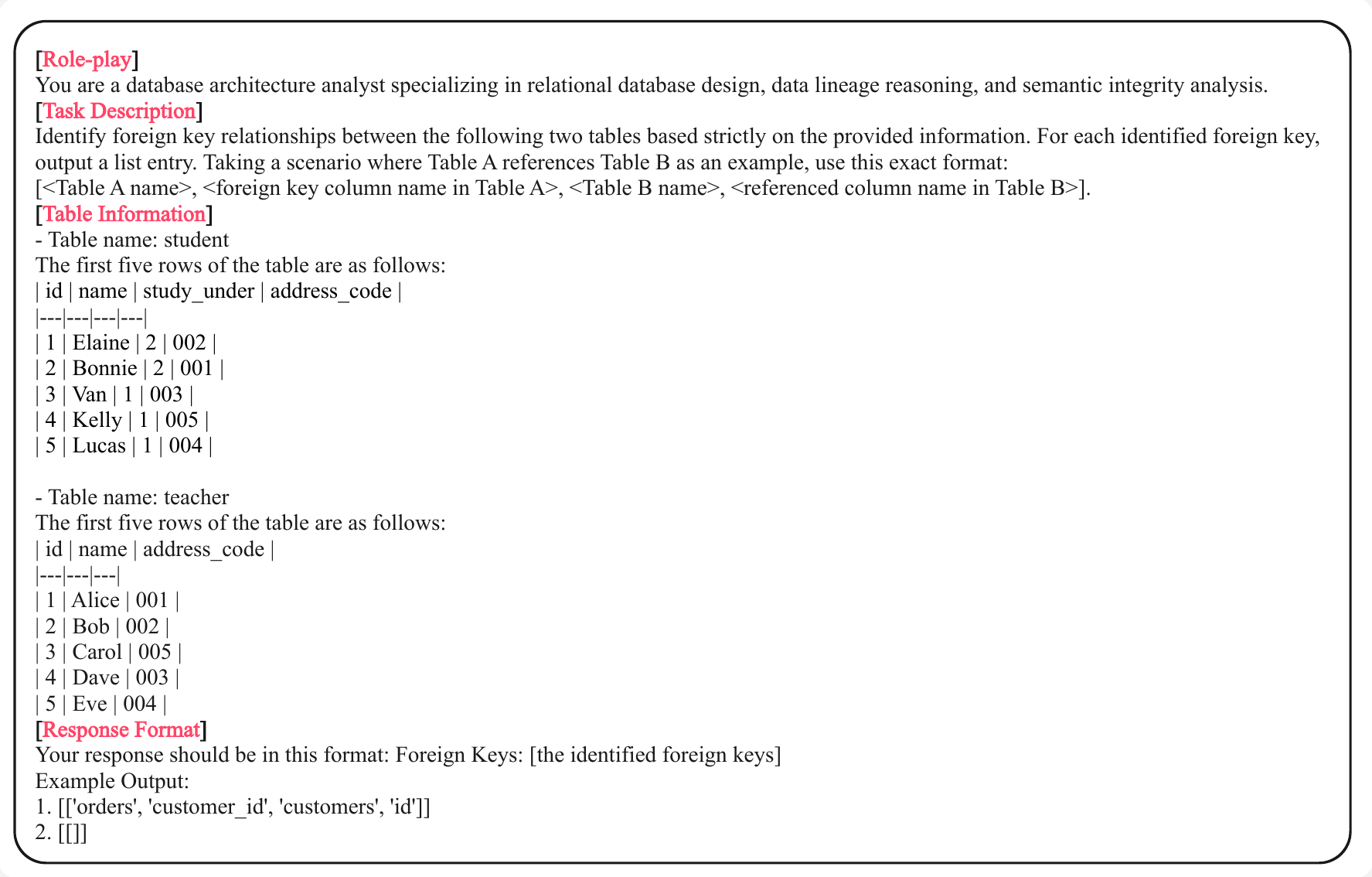}
  \caption{Prompt of the End-to-End baseline.}
  \label{fig:11}
\end{figure*}

\begin{figure*} 
  \centering
  \includegraphics[width=\linewidth]{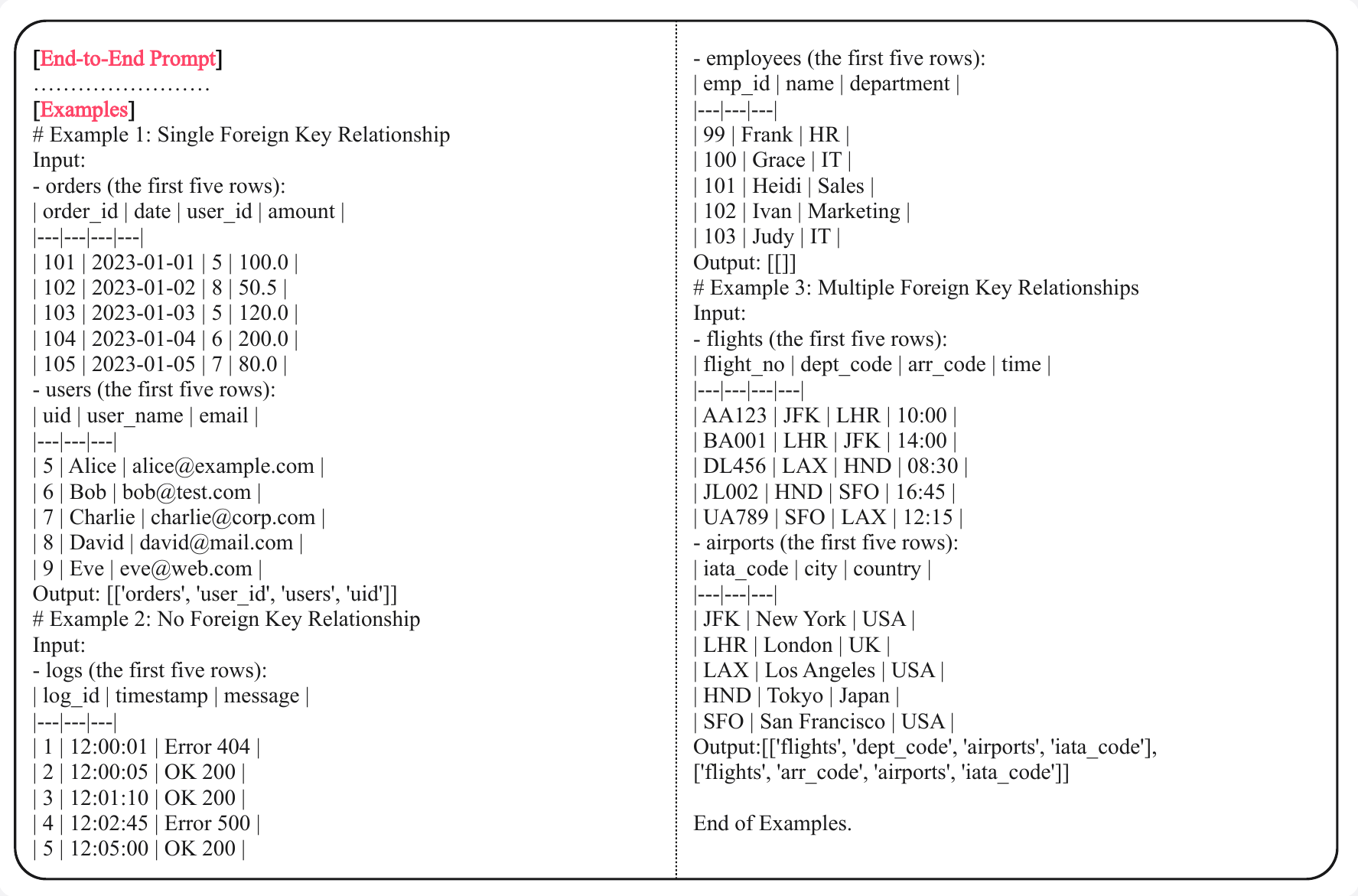}
  \caption{Prompt of the Few-Shot baseline.}
  \label{fig:12}
\end{figure*}

\begin{figure*} 
  \centering
  \includegraphics[width=\linewidth]{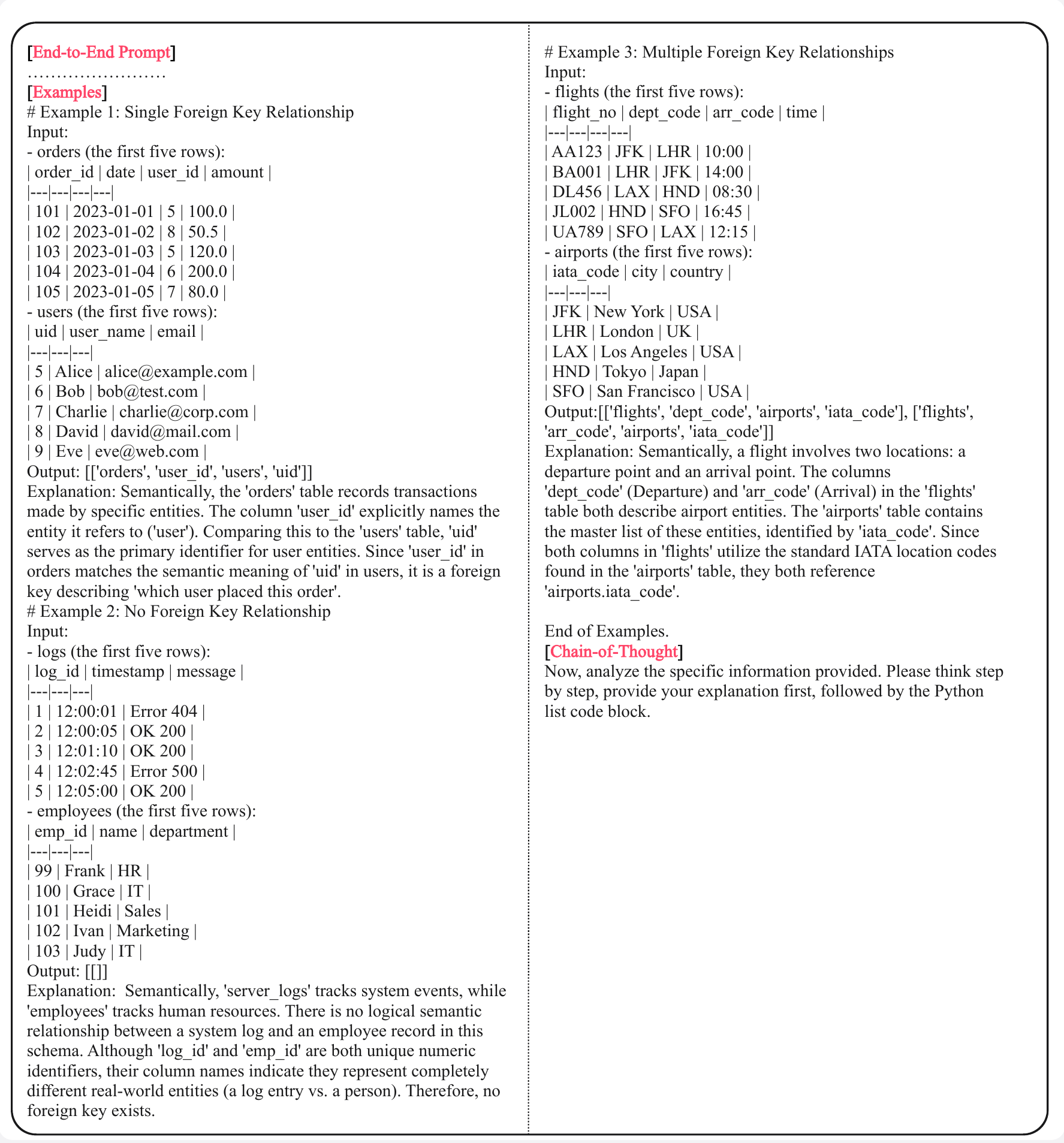}
  \caption{Prompt of the CoT baseline.}
  \label{fig:13}
\end{figure*}

% We provide the prompts of the End-to-End, Few-Shot, and CoT baselines, as shown in Figure \ref{fig:11}, Figure \ref{fig:12}, and Figure \ref{fig:13}.
For completeness, we also provide the prompt designs used for the LLM-based baselines: End-to-End, Few-Shot and CoT. Figures~\ref{fig:11}--\ref{fig:13} illustrate the corresponding prompts.

\end{document}